\documentclass[amsmath,amssymb,aps,pra,preprint,groupedaddress]{revtex4-1}
\usepackage[dvipdfmx]{graphicx}% Include figure files
\usepackage{dcolumn}% Align table columns on decimal point
\usepackage{bm}% bold math
\usepackage{longtable}
\usepackage{ulem}
\usepackage{xcolor,soul}

\begin{document}
	% Use the \preprint command to place your local institutional report
	% number in the upper righthand corner of the title page in preprint mode.
	% Multiple \preprint commands are allowed.
	% Use the 'preprintnumbers' class option to override journal defaults
	% to display numbers if necessary
	%\preprint{}

\title{Monte Carlo Studies on Geometrically Confined Skyrmions in Nanodots: Stability and Morphology under Radial Stresses}% Force line breaks with \\

\author{G. Diguet}
\email{gildas.diguet.d4@tohoku.ac.jp}
\affiliation{%
	Micro System Integration Center, Tohoku University, Sendai, Japan% Force line breaks with \\ if necessary
}%
\author{B. Ducharne}
% \email{third.author@anotherinstitution.edu}
\affiliation{%
	INSA Lyon, Universite de Lyon, Villeurbanne Cedex, France% Force line breaks with \\ if necessary
}%
\affiliation{ELyTMaX, CNRS-Universite de Lyon-Tohoku University, Sendai, Japan}
\author{S. El Hog}
% \email{third.author@anotherinstitution.edu}
\affiliation{%
	Universit$\acute{\rm e}$ de Monastir (LMCN), Monastir, Tunisie% Force line breaks with \\ if necessary
}
\author{F. Kato}%
\affiliation{
	National Institute of Technology (KOSEN), Ibaraki College, Hitachinaka, Japan.% Force line breaks with \\ if necessary
}
\author{H. Koibuchi} % Write as First name Surname
% \email[Corresponding author: ]{koibuchi@gm.ibaraki-ct.ac.jp; koibuchih@gmail.com}
\email{koi-hiro@sendai-nct.ac.jp; koibuchih@gmail.com}
\affiliation{
	National Institute of Technology (KOSEN), Ibaraki College, Hitachinaka, Japan.% Force line breaks with \\ if necessary
}
\author{T. Uchimoto}
\affiliation{%
	Institute of Fluid Science (IFS), Tohoku University, Sendai, Japan% Force line breaks with \\ if necessary
}%
\affiliation{ELyTMaX, CNRS-Universite de Lyon-Tohoku University, Sendai, Japan}
\author{H. T. Diep}
\email{diep@cyu.fr}
\affiliation{%
	CY Cergy Paris University, Cergy-Pontoise, France% Force line breaks with \\ if necessary
}%

\begin{abstract}
We numerically study the stability and morphology of geometrically confined skyrmions in nanodots using Finsler geometry (FG) modeling technique. The FG model dynamically implements anisotropies in ferromagnetic interaction, Dzyaloshinskii-Moriya interaction, and magneto-elastic coupling in response to mechanical stresses. Without the stresses, there exists a geometrically confined effect originating from the surface effect of small nanodots, in which skyrmions are stabilized under a low external magnetic field. This surface effect is enhanced by radial stresses, which significantly reduce the surface DMI compared to the bulk DMI. The radial stresses also alter the interactions to be anisotropic. Owing to these position- and direction-dependent interactions, incomplete skyrmions emerge at the center of the nanodots under the tensile stress. In addition to the incomplete skyrmions, target skyrmions are observed under the compressive stress.  Our numerical results indicate that the strain-enhanced surface effect and the strain-induced interaction anisotropies suitably explain the skyrmion stability in nanodots with zero magnetic field.
\end{abstract}

% Keywords: Skyrmions; Stability; Geometric confinement; Radial stress; Monte Carlo; 

\maketitle
% Section
\section{Introduction\label{intro}}
% stability / instability
% G.C.
% Magnetic anisotropy
% 
A topologically stable spin configuration called skyrmion emerges in chiral magnetic materials under competition between ferromagnetic interaction (FMI) and Dzyaloshinskii-Moriya interaction (DMI), and many studies have been conducted \cite{Romming-etal-Science2013,Fert-etal-NatReview2017,Zhang-JPCM2020,Tokura-Kanazawa-ACS2020,Gobel-etal-PhysRep2021,Wang-etal-JMMM2022,Li-etal-Wiley2022} since its experimental discovery \cite{Uchida-etal-SCI2006,Pfleiderer-etal-Science2009,Yu-etal-Nature2010}. It is well known that skyrmions are stabilized by an external magnetic field in Zeeman energy and mechanical strains from magnetoelastic coupling (MEC) and magnetic anisotropy \cite{Bogdanov-PRL2001,Butenko-etal-PRB2010}.  The confinement of skyrmions, called geometric confinement (GC), has also been proposed as a stabilization technique and emphasized to be caused by surface or edge boundary conditions for spin configurations in nanodot \cite{Rohart-Thiaville-PRB2013}, and several experimental and numerical studies on GC and strain effects have been conducted \cite{HDu-etal-NatCom2015,Matsumoto-etal-NanoLett2018,CJin-etal-NatCom2017,ZHou-etal-AcsNano2019,PHo-etal-PRAp2019,Seki-etal-PRB2017,YWang-etal-NatCom2020,Li-etal-PRAP2020}. 

As a GC effect inside nanowires, thermodynamically stable unusual skyrmions were theoretically investigated \cite{Leonov-etal-EPJConf2014}. One is called "target skyrmion”  surrounded by multiple circular stripes. The target skyrmions can be identified with "helix" numerically investigated in the presence of dipolar coupling \cite{Du-etal-PRBL2013} and were experimentally observed in nanodisks of FeGe  \cite{Zheng-etal-PRL2017} and Permalloy \cite{Kent-etal-APL2019}. The other is of a vortex-like structure called "incomplete skyrmions" with the topological charge or skyrmion number $N_{\rm sk}$ less than 1. The incomplete skyrmions were numerically studied for their responses to the variations of external magnetic field and nanodot diameter in the ground states \cite{Beg-etal-SRep2015,Beg-etal-PRB2017}. 

Y. Wang et al. demonstrated the application of GC of skyrmions to electric switching and confirmed a controllability of skyrmion on nanodots of diameter 350 nm without a magnetic field, in combination with GC and strain effects \cite{YWang-etal-NatCom2020}. Their results show that the strain effects on magnetic anisotropy and DMI are crucial for the morphological changes. The skyrmion stabilization was numerically confirmed on nanodots without an external magnetic field under a specific boundary condition \cite{Li-etal-PRAP2020}, and the effects of magnetic anisotropy were studied numerically \cite{Mehmood-etal-JMMM2022}. A GC effect for stabilizing skyrmion vortex  was also observed in tetrahedral nanoparticles \cite{Niitsu-etal-Natmatt2022}. However, the origins of GC and the effects of strain are unclear and need to be studied further.

In this study, we apply Finsler geometry (FG) modeling to clarify the strain effect on morphological changes in spin configurations in nanodots using the Monte Carlo (MC) simulation technique \cite{Diguet-etal-ICFD2023}. Dynamical anisotropies in the FMI, DMI, and MEC emerging under radial stresses are numerically evaluated to explain the skyrmion stability. For the stabilization, FG modeling implements anisotropic effects of strains on interactions by internal degrees of freedom, such as the strain direction, which dynamically modifies the interactions to be direction-dependent and position-dependent.  For this simple geometric treatment of interactions, FG modeling allows us to perform theoretical studies on skyrmions based on the same input-output relation as experimental studies, leading to a better understanding of the surface effect of skyrmions in nanodots. This is the main advantage of FG modeling over standard modeling, in which interactions are manually controlled by fixing the coefficients to be anisotropic.

This study is motivated by the experimental study in Ref. [20]. However, the targeted material is simplified for modeling: for example a spontaneous in-plane magnetic anisotropy assumed in \cite{YWang-etal-NatCom2020} is not necessary in our model. Furthermore, our model is based on a single layer that may not be as thin as experimental samples,  and the experimental technique for the application of radial strains to nanodots is not the same as in our model. Nevertheless, we believe that these differences do not change the main effects of the radial strains observed in \cite{YWang-etal-NatCom2020} and the present numerical study.

% Section
\section{3D cylindrical lattices and Hamiltonian \label{Hamiltonian-lattice}}
% subsection
\subsection{3D cylindrical lattices and radial stresses\label{lattices}}
% f-1
\begin{figure}[h]
	\centering{}\includegraphics[width=10.5cm,clip]{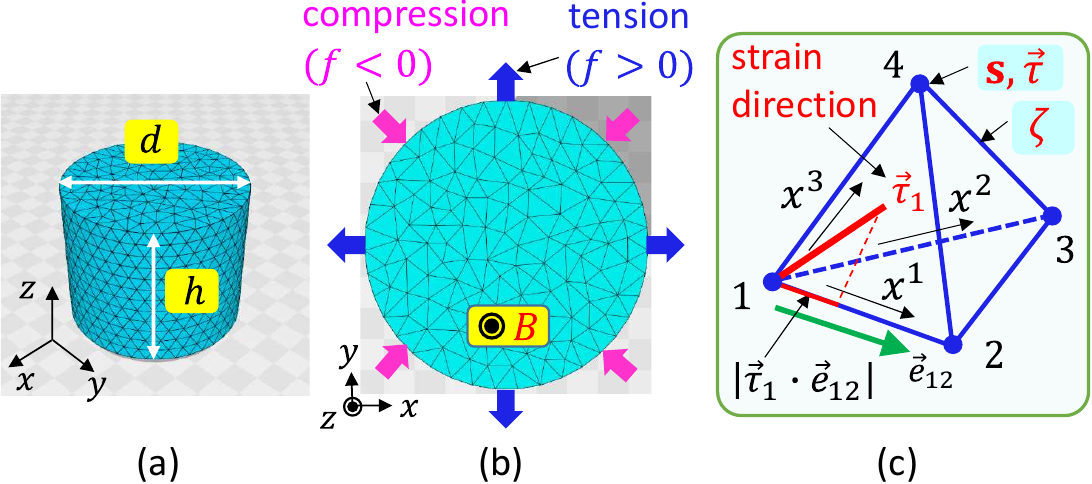}
	\caption{
		(a) A cylindrical lattice of radius $d$ and  height $h$, (b) illustrations of tensile ($f\!>\!0$) and compressive ($f\!<\!0$) stresses radially applied to cylindrical lattice, (c) strain direction $\vec{\tau}_1$ at vertex 1  and its component $|\vec{\tau}_1\cdot\vec{e}_{12}|$, where $\|\vec{e}_{12}\|\!=\!1$, along a local coordinate axis $x^1$ of a tetrahedron with vertices 1, 2, 3 and 4. The variables ${\bf s}, \vec{\tau}$ are defined at vertices and $\zeta$ on bonds, and these are introduced in the following subsection.  Free boundary condition is assumed for all surfaces.		\label{fig-1} }
\end{figure}
Cylindrical lattices constructed by the Voronoi tessellation technique \cite{Friedberg-Ren-NPB1984} are used to simulate skyrmions in nanodots \cite{YWang-etal-NatCom2020} (Fig. \ref{fig-1}(a)). The total number of vertices is $N\!=\!1533$, 2083, 5430, 8465, and 11932, and the ratio  $R\!=\!d/h$ of diameter $d$ and a fixed height $h$ is approximately given by  $R\!=\!1.01$, $1.20$, $2.00$, $2.51$, and $R\!=\!3.00$ (see Appendix \ref{App-A}  for further details of the lattice structure).  Mechanical stresses, $f\!>\!0$ and $f\!<\!0$ applied along the radial direction  (Fig. \ref{fig-1}(b)), cause a variation of strain variable $\vec{\tau}$  (Fig. \ref{fig-1}(c)) leaving the lattice structure non-deformed, and a magnetic field $B$ is applied along the $z$ direction. The direction of magnetic field and that of stresses are the same as those in  \cite{YWang-etal-NatCom2020}. A free boundary condition is assumed for all surfaces, including the lower disk, which contacts with a polymer substrate for strain transfer in the reported experiment  in \cite{YWang-etal-NatCom2020}; hence, the boundary condition in the present study is not exactly the same as that of the experiment.

% subsection
\subsection{Hamiltonian and Monte Carlo\label{Hamiltonian}}
% f-2
\begin{figure}[h]
	\centering{}\includegraphics[width=11.5cm,clip]{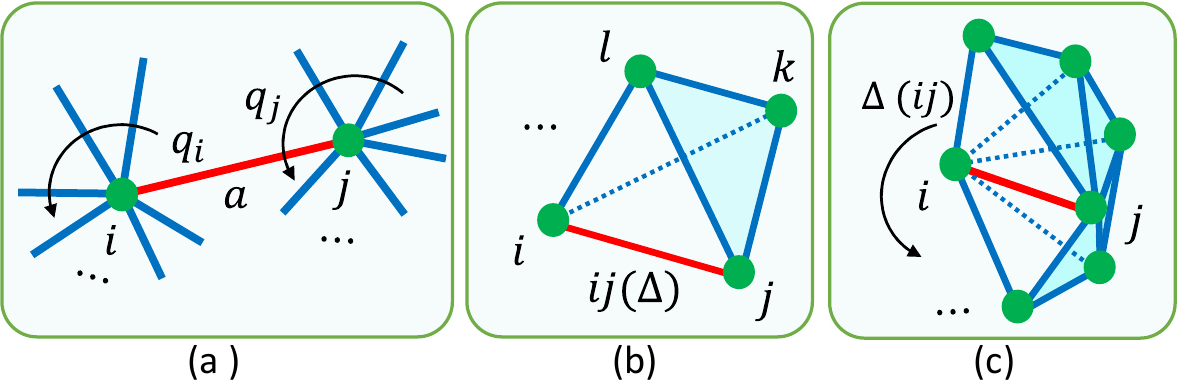}
	\caption{
		(a) Illustration of bond $a$ or $ij$ and the connected bonds. The total number of the connected  bonds at vertex $i$ ($j$) is $q_i\!-\!1$ ($q_j\!-\!1$), where $q_i$ ($q_j$) is the coordination number at vertex $i$ ($j$). Illustrations of (b) $\sum_{ij(\Delta)}$; the sum over bonds $ij(\Delta)$ of tetrahedron $\Delta$, and (c) $\sum_{\Delta(ij)}$; the sum over tetrahedrons $\Delta(ij)$ sharing bond $ij$. $\sum_{ij(\Delta)}1\!=\!6$ for all $\Delta$, while the number  $n_{ij}\!=\!\sum_{\Delta(ij)}1$ depends on bond $ij$, and we have the identity $6N_{\rm tet}\!=\!\sum_\Delta \sum_{ij(\Delta)}1\!=\!\sum_{ij}\sum_{\Delta(ij)}1\!=\!\sum_{ij}n_{ij}$, where $N_{\rm tet}$ is the total number of tetrahedrons. \label{fig-2} }
\end{figure}
The discrete Hamiltonian is 
\begin{eqnarray}
	\label{total-Hamiltonian}
	H({\bf s},\vec{\tau}, \zeta)=\lambda H_{\rm FM}+DH_{\rm DM}-H_B-\alpha H_{\rm ME}-H_f-\delta H_{\zeta},
\end{eqnarray}
where ${\bf s} (\in S^2: {\rm unit\; sphere})$ and $\vec{\tau} (\in S^2/2: {\rm half\; sphere})$ denote the spin and strain variables, respectively, defined at every lattice vertex, and $\zeta (\in \{1,-1\})$  on the bonds (Fig. \ref{fig-1}(c));  $\zeta$ is a variable to implement a dynamical anisotropy in MEC. The symbols $\lambda, D, \alpha, \delta$ on the right-hand side are the interaction coefficients; $\lambda$, $D$ and $\alpha$ are strengths of FMI,   DMI  and MEC, respectively, and $\delta$  denotes a strength of $\zeta$ interaction for Finsler length of MEC (Appendix \ref{App-B}).

The terms on the right hand side are given as follows:
\begin{eqnarray}
	\label{Hamiltonians}
	\begin{split}
		& H_{\rm FM}=\sum_{\Delta} \sum_{ij(\Delta)}\Gamma_{ij}(\vec{\tau})\left(1-{\bf s}_i\cdot{\bf s}_j\right),\quad
		H_{\rm DM}=\sum_{\Delta} \sum_{ij(\Delta)}\Gamma_{ij}(\vec{\tau})\vec{e}_{ij}\cdot {\bf s}_i\times{\bf s}_j,\\
		& H_B=\sum_i {\bf s}_i\cdot\vec{B},\; \vec{B}=(0,0,B), \quad  
		 H_{\rm ME}=f\sum_{\Delta} \sum_{ij(\Delta)}\Omega_{ij}(\vec{\tau},\zeta)\left({\bf s}_i\cdot{\bf s}_j\right)^2,\\
		& H_f={\rm sgn}(f)\sum_{i} \left(\vec{\tau}_i\cdot \vec{f}\right)^2,\quad \vec{f}={f}\vec{e}^{\; r},  \quad {\rm sgn}(f)=\left\{ \begin{array}{@{\,}ll}
			1\quad  (f>0) \\
			-1\quad  (f<0)
		\end{array} 
		\right., \\
		& H_\zeta=\sum_{(ab)}\zeta_a \zeta_b
	\end{split}
\end{eqnarray}
The first term $H_{\rm FM}$ is an FMI energy that is modified to have a $\vec{\tau}$-dependent interaction coefficient $\Gamma_{ij}(\vec{\tau})$, which depends on a small number $v_0$ playing a role in the strength of anisotropy (see Appendix \ref{App-B} for FG modeling details). The same $\Gamma_{ij}(\vec{\tau})$ is included in the second term $H_{\rm DM}$ for DMI energy. The third term $H_B$ is the Zeeman energy, and the fourth term $H_{\rm ME}$ is the energy for the ME coupling quadratic with respect to ${\bf s}$ with a coefficient $\Omega_{ij}(\vec{\tau},\zeta)$, which is dynamically variable depending on $\zeta$ and, in this sense, different from $\Gamma_{ij}(\vec{\tau})$, which varies depending only on $\vec{\tau}$,  in $H_{\rm FM}$ and $H_{\rm DM}$ (Appendix \ref{App-B}). The fifth term $H_f$ denotes the energy required for the response of strain $\vec{\tau}$ to an external stress $\vec{f}$ along the radial direction $\vec{e}^{\; r}$.  The final term $H_\zeta$, which is not included in the model in Ref. \cite{Diguet-etal-ICFD2023}, describes the dynamical Finsler length for an anisotropic MEC in $H_{\rm ME}$.  $H_\zeta$ is defined as the sum of connected bonds $a$ and $b$ denoted by $(ab)$ (Fig. \ref{fig-2}(a)). We note that $H_\zeta$ is equivalent to the Potts model Hamiltonian, $H_{\rm P}\!=\!\sum_{ab} \delta_{\zeta_a,\zeta_b}$, where $\delta_{\zeta_a,\zeta_b}\!=\!1\; (\zeta_a\!=\!\zeta_b)$ and $\delta_{\zeta_a,\zeta_b}\!=\!0\; (\zeta_a\!\not=\!\zeta_b)$. $H_{\rm FM}$,  $H_{\rm DM}$  and  $H_{\rm ME}$  are commonly defined by  $\sum_{\Delta} \sum_{ij(\Delta)}$, the sum over tetrahedrons $\Delta$, where $\sum_{ij(\Delta)}$ denotes the sum over bonds $ij(\Delta)$ of tetrahedron $\Delta$  (Fig. \ref{fig-2}(b)).

In the case of a zero stress $f\!=\!0$, the spin configurations are determined only by the first three terms. For this reason, no spontaneous magnetic anisotropy is assumed in the model; therefore, the material studied in the present paper is not exactly the same as that experimentally studied in Ref. \cite{YWang-etal-NatCom2020} in addition to a difference in the application technique of radial stress on nanodots. Moreover, the materials studied in Ref. \cite{YWang-etal-NatCom2020} are layered ones and a Neel-type skyrmion is assumed in their simulations; in contrast, the materials in the present paper are single-layer and not always thin, in which a Bloch-type skyrmion is assumed.

The partition function to be simulated by MC technique is given by 
\begin{eqnarray}
	\label{part-func}
	Z=\sum_{{\bf s}}\sum_{\vec{\tau}}\sum_\zeta \exp \left[-H({\bf s},\vec{\tau},\zeta)/T\right],
\end{eqnarray}
where $T$ is the temperature and $\sum_{s,\tau,\zeta}$ denotes the sum of all possible configurations of ${\bf s},\vec{\tau}$ and $\zeta$. The partition cannot be numerically obtained, and instead, the sum over configurations, defined by $\sum_{{\bf s}}\sum_{\vec{\tau}}\!=\! (\int\Pi_{i=1}^N d\vec{s}_i)(\int\Pi_{i=1}^N d\vec{\tau}_i)$ the multiple integrals on the unit sphere and $\sum_\zeta\!=\!\Pi_{i=1}^N\sum_{\zeta_i\in \{-1,1\}}$, is simulated by  Metropolis importance MC sampling \cite{Metropolis-JCP-1953,Landau-PRB1976}, where $\vec{\tau}_i$ is identified with $-\vec{\tau}_i$ in the Hamiltonians. The mean value of physical quantity $P_q$ is defined by the ensemble average $\langle P_q\rangle\!=\!\sum_{s,\tau,\zeta}P_q \exp (-H/T)/Z$, which is calculated using Metropolis MC technique \cite{Metropolis-JCP-1953,Landau-PRB1976} with the sample average $\langle P_q\rangle\!=\!\sum_i P_q(i)/\sum_i 1$, where $P_q(i)$ is $i$-th sample evaluated by the lattice average.  A new variable ${\bf s}^{\prime}$ is randomly generated, independent of the present ${\bf s}$ and updated with probability ${\rm Min}[1,\exp -(H_{\rm new}\!-\!H_{\rm old})]$. The same procedure is applied to the variable $\vec{\tau}$. When updating $\zeta (\in \!\{-1,1\})$, we fix $\zeta^{\prime}$ to the opposite to the present $\zeta$ and accept  $\zeta^{\prime}$ with the same procedure. $N$ consecutive updates for each of the three variables are examined, and a uniform random number is used to select $N$ from $N_B$ $\zeta$-variables defined on the bonds, where $N_B$ is the total number of bonds (Appendix \ref{App-A}).

The total number of MC sweeps (MCS) for the calculation of physical quantities is $5.25\!\times\! 10^7$ for the lattices of $N\!\leq\!8465$ and $(1.75\sim3.5)\!\times\! 10^8$  for the $N\!=\!11932$ lattice.  The samples are calculated for every 500 MCS after the thermalization iterations of $7.5\!\times\! 10^6$ $\sim$  $12.5\!\times\! 10^6$ MCS for the lattices of $N\!\leq\!8465$ and $5\!\times\! 10^7$ MCS for the $N\!=\!11932$ lattice starting with random initial configurations.

% Section
\section{Results \label{results}}
% subsection
\subsection{Size effect without strains \label{size_effect}}
% f-3
\begin{figure}[h]
	\centering{}\includegraphics[width=11.5cm,clip]{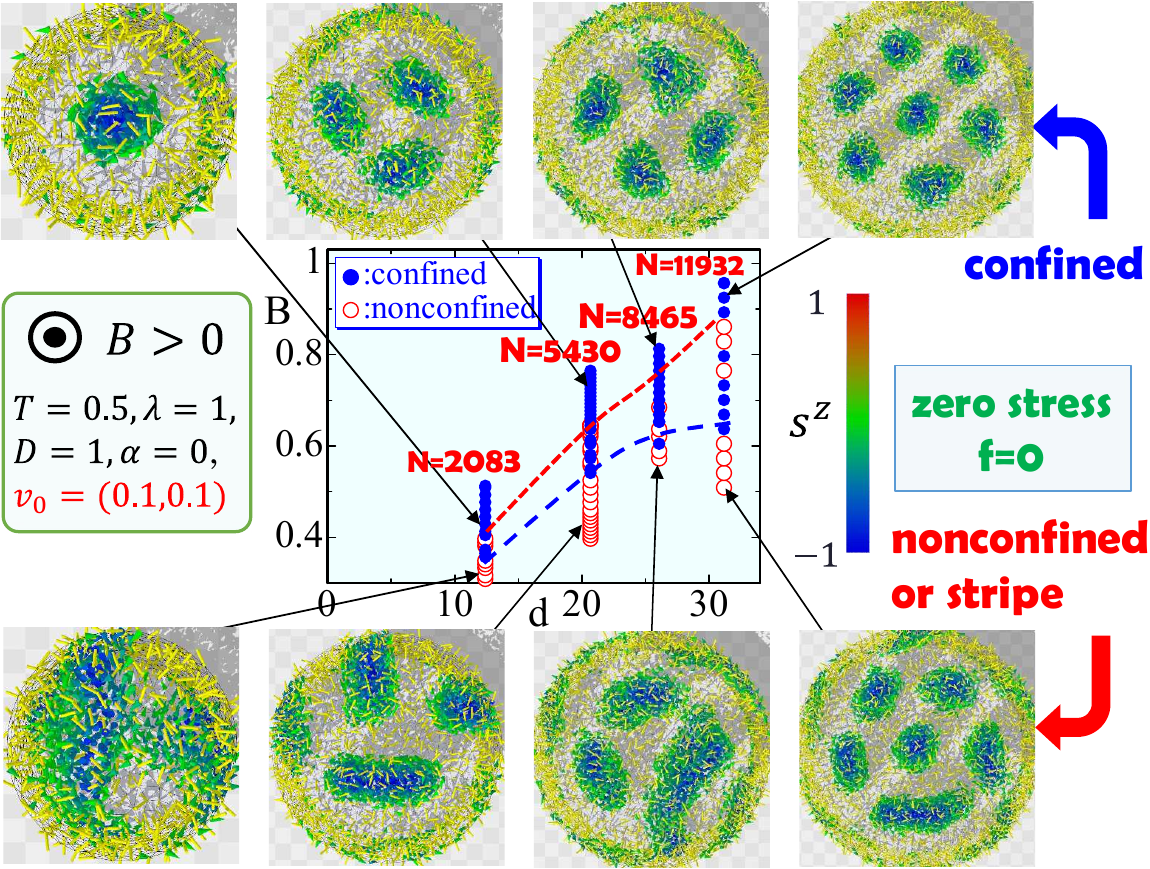}
	\caption{ Phase diagram of $B$ vs.  $d$ for confined skyrmion (\textcolor{blue}{$\bullet$}) and non-confined or stripe (\textcolor{red}{$\circ$}) observed on four different lattices, where $d$ is the diameter of lattice in the unit of the edge length ($a\!=\!1$) of a regular triangle (Appendix \ref{App-A}). Spins of $s^z\!<\!0$ are plotted with small cones in the snapshots. Small cylinders or lines, shown only on the surface for clear visualization, denote $\vec{\tau}$, which is isotropic and does not influence the spins. 
				\label{fig-3} }
\end{figure}
First, we demonstrate a size effect that enables skyrmions to emerge on smaller nanodots with a small external magnetic field. The assumed parameters are shown in Fig.  \ref{fig-3}. The temperature is fixed to $T\!=\!0.5$ in all simulations, because it is low enough to see skyrmions. $v_0\!=\!(0.1,0.1)$ denotes that $v_0\!=\!0.1$ for both the FMI and DMI in Eq. (\ref{Finsler-unit-length-FDMI}). $v_0$ plays a role in the strength of anisotropy in the cases that strains $\vec{\tau}$ are uniformly aligned; however, it does not play any role in anisotropy of interactions in the case of $f\!=\!0$, where $\vec{\tau}$ is isotropic.  Plots in Fig. \ref{fig-3} represent configurations of confined skyrmion (\textcolor{blue}{$\bullet$}) and non-confined or stripe (\textcolor{red}{$\circ$}) observed on four different lattices. The color code represents $s^z$ and the spins of $s^z\!<\!0$ are plotted with cones, and small cylinders or lines represent the direction of $\vec{\tau}$.

All the plotted data are obtained by visually evaluating whether the skyrmions are confined using snapshots. The "non-confined" means (i) stripe configurations or (ii) skyrmions are located at the boundary, and the "confined" means (iii) skyrmions are inside the boundary. However, these criteria are not strict, and the phase diagram shown in Fig. \ref{fig-3} is a rough estimate. The dashed line in red (blue) shows the upper (lower) limit of $B$ for the non-confined (confined) phase. The region between these two dashed lines is similar to the range of two-phase coexistence; however, it does not always imply a first-order phase transition because no clear jump is observed in the physical quantities, including the topological charge between the confined skyrmion and non-confined skyrmion phases. Interestingly, the decrease of $B$ for the confined region with decreasing $d$ is qualitatively consistent with experimental data reported in Ref. \cite{YWang-etal-NatCom2020}.

Notably, the confinement mechanism in the model originates from a surface effect, in which the absolute value of the surface DMI coefficient is smaller than that of the bulk DMI \cite{Diguet-etal-JMMM2023}. For this reason, skyrmions hardly appear on the periphery region. 
 A static part of this difference in the DMI strength for $f\!=\!0$ is intuitively understood from the discrete form of $H_{\rm DM}$ in Eq. (\ref{Hamiltonians}) (Appendix \ref{App-B}), and a dynamical enhancement of the difference for $f\!>\!0$ will be shown below.

% subsection
\subsection{Strain effect \label{strain_effect}}
% f-4
\begin{figure}[h]
	\centering{}\includegraphics[width=11.5cm,clip]{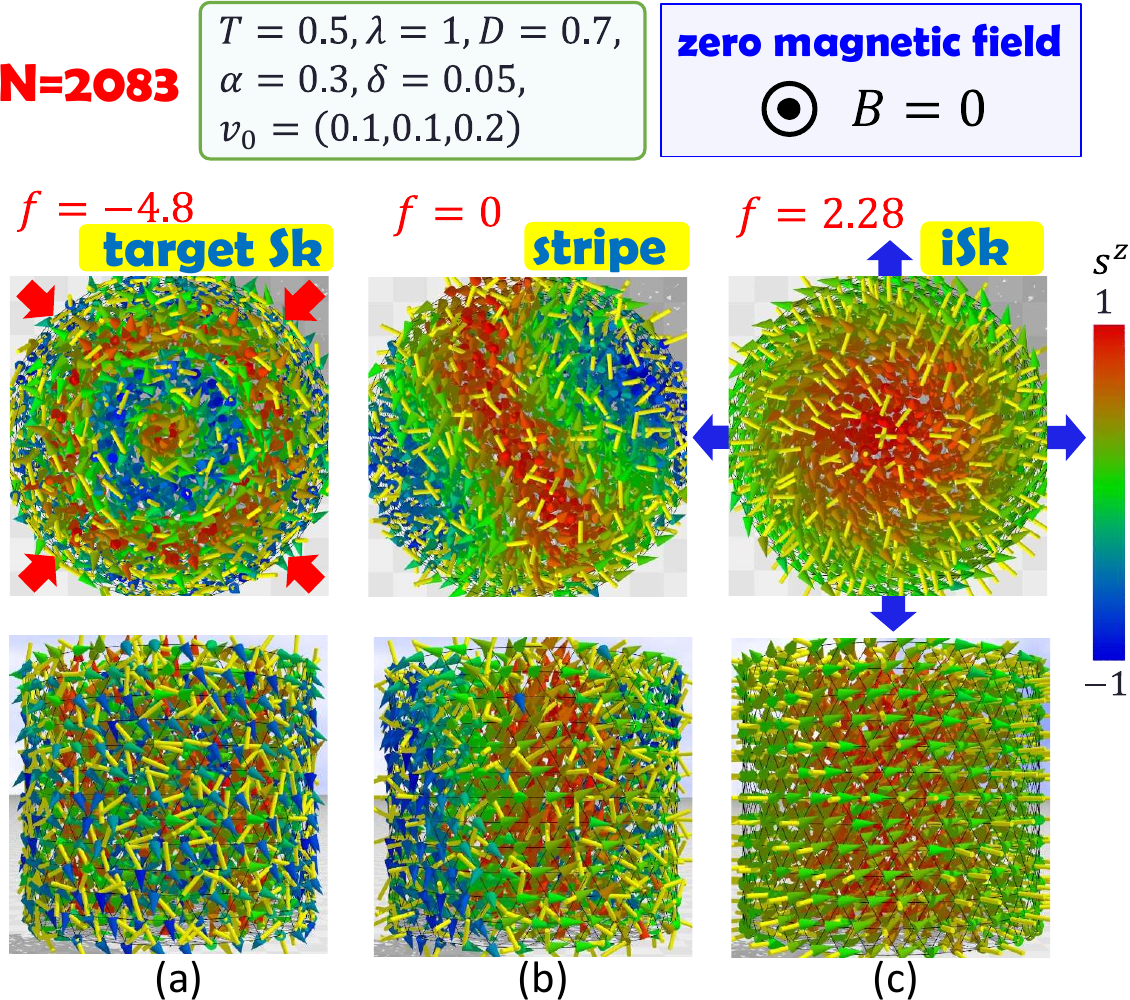}
	\caption{
		The snapshots obtained for (a) $f\!=\!-4.8$ (compression), (b) $f\!=\!0$ , and (c) $f\!=\!2.28$ (tension) on a small lattice of size $N\!=\!2803$ for clear visualization. We find the target skyrmions (target Sk), stripes, and incomplete skyrmion (iSk) phases for negative, zero, and positive stresses. These three phases correspond to vortex,  stripe and skyrmion in the thin-layered case of the reported experimental data \cite{YWang-etal-NatCom2020}. Small cylinders for $\vec {\tau}$ are shown only on the surface for clear visualization. The parameters $T,\lambda, D, \alpha, \delta$ and $v_0$ shown in the figure are used in all simulations below. \label{fig-4} }
\end{figure}
Next, we show the effects of the strain caused by radial tension ($f>0$) and compression ($f<0$) under zero external magnetic field ($B\!=\!0$). To visualize the configurations of ${\bf s}$ and $\vec{\tau}$, a small cylinder of size $N\!=\!2803$ is used in this subsection.  The assumed parameters are shown in Fig.  \ref{fig-4}; $v_0\!=\!(0.1,0.1,0.2)$ denotes that $v_0\!=\!0.1$ for both FMI and DMI in Eq. (\ref{Finsler-unit-length-FDMI}) and  $v_0\!=\!0.2$ for MEC in Eq. (\ref{Finsler-unit-length-ME}), which are fixed in this study.

Snapshots are shown in Figs. \ref{fig-4}(a)--(c), where top and side views are plotted in (a) $f\!=\!-4.8$, (b) $f\!=\!0$, and (c) $f\!=\!2.28$. Small cylinders representing $\vec{\tau}$ can be viewed as (a) a spiral along the $z$ direction, (b) isotropic, and (c) radially aligned. These observations are consistent with expectations because $\vec{\tau}$ represents the elongation direction owing to the radial stress $f$. 

The configuration shown in Fig. \ref{fig-4}(a) is a skyrmion enclosed by circular stripes corresponding to the target skyrmion denoted by "target Sk" or simply by "target" \cite{Leonov-etal-EPJConf2014}, which is not exactly the same as "vortex" observed in Ref. \cite{YWang-etal-NatCom2020} for the thin-layered case. The configuration in Fig. \ref{fig-4}(c) is also a skyrmion, which is called "incomplete skyrmion" in \cite{Leonov-etal-EPJConf2014} denoted by iSK in the figure. The direction of ${\bf s}$ at the periphery is not completely opposite to that at the center \cite{Leonov-etal-EPJConf2014,Beg-etal-SRep2015,Beg-etal-PRB2017}. However, in their experiment on the thick-layered case, stripes were observed, which are consistent with our numerical data. We checked that these morphological changes under radial stress can also be observed in Neel-type skyrmions. Thus, we consider that these morphological changes under radial stresses correspond to vortices, stripes, and skyrmions in the thin-layered case of reported experimental data \cite{YWang-etal-NatCom2020}, although the vortex and skyrmions in \cite{YWang-etal-NatCom2020} are not exactly corresponding to the target and incomplete skyrmions in this paper, respectively.

It should be noted that the spin direction at the center of the target skyrmion shown in Fig. \ref{fig-4}(a) is spontaneously determined and opposite to that shown in Fig. \ref{fig-3}. This spontaneous direction occurs because $B\!=\!0$, in contrast with the case in Ref. \cite{YWang-etal-NatCom2020}, where a small magnetic field is applied for atomic microscopy measurements, even with zero external magnetic field. For the same reason, the polarity at the center of the incomplete skyrmion in Fig. \ref{fig-4}(c) is also spontaneously determined. Note also that the radial stress, for example, compression, is well defined only in 3-dimensional models because of the induced deformation of the nanodots along the $z$ direction indicated by the $\vec{\tau}$ direction.

% f-5
\begin{figure}[h]
	\centering{}\includegraphics[width=11.5cm,clip]{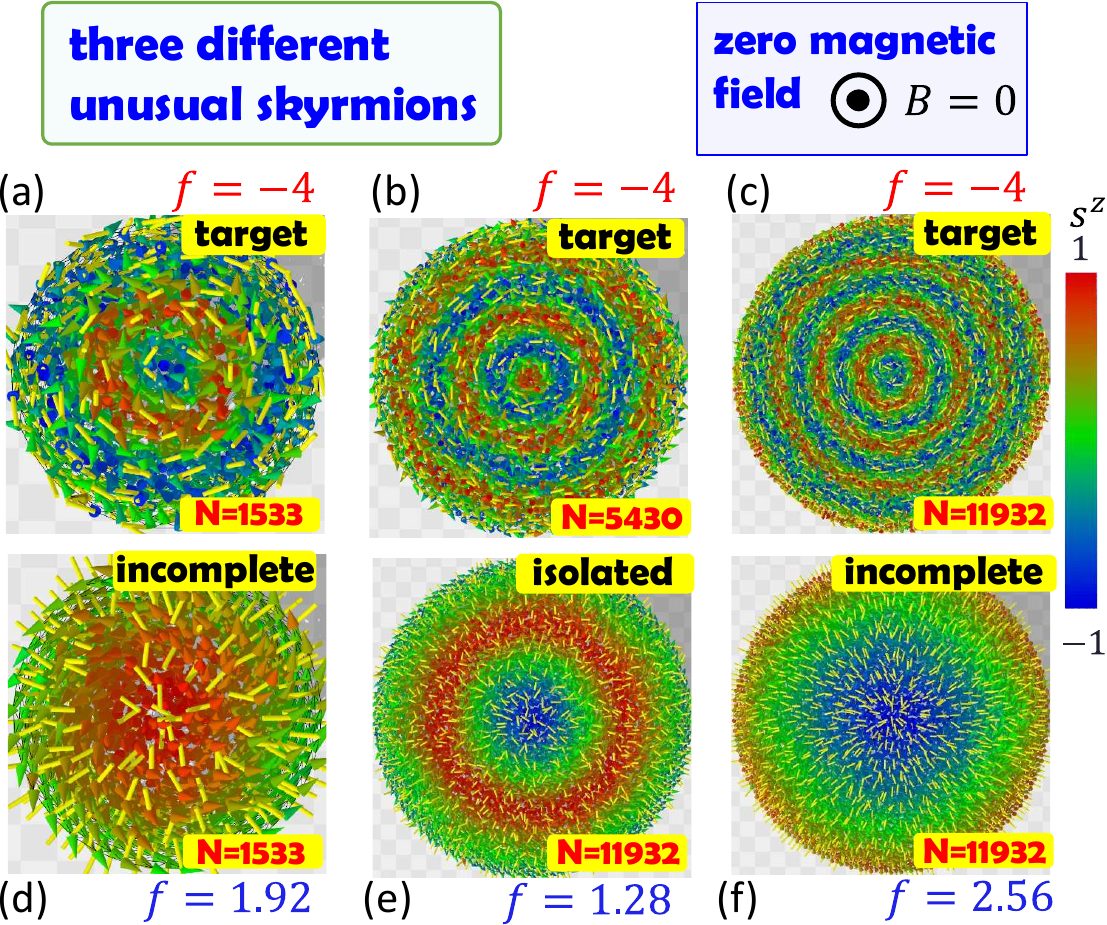}
	\caption{
		The snapshots of target skyrmions on the lattices of size (a) $N\!=\!1533$, (b) $N\!=\!5430$ and (c) $N\!=\!11932$, and (d) incomplete skrymion on the $N\!=\!1533$ lattice, (e) an isolated skrymion and (f) incomplete skrymion on the $N\!=\!11932$ lattice. The polarity variation between the center and periphery is less than $\pi$ in (d), (f), whereas it is more than $\pi$ in (e).   \label{fig-5} }
\end{figure}

To observe the lattice size dependence, snapshots of the target, incomplete, and isolated skyrmions are shown in Figs. \ref{fig-5}(a)--(f) for lattices of size $N\!=\!1533$, $N\!=\!5430$ and $N\!=\!11932$. Note that the total number of circular stripes of $N\!=\!5430$ and $N\!=\!11932$ lattices is larger than that of the $N\!=\!1533$ lattice  because the lattice diameter $d$ approximately increases $1.97$ and $2.97$ times, respectively (Table \ref{table-1} in Appendix \ref{App-A}). Such a change in the number of morphology can also be observed in the stripe phase. Therefore, the number of circular stripes  is determined by the scale $\lambda/D$ of the FMI and DMI coefficients in Eq. (\ref{total-Hamiltonian}).

However, the total number (=1) of incomplete skyrmions remains unchanged and is independent of $d$ on the lattices of $N\!=\!1533$ in Fig. \ref{fig-5}(d) and $N\!=\!5430$ (not shown in Fig. \ref{fig-5}), in sharp contrast to the case of nonzero magnetic field with $f\!=\!0$  in Fig. \ref{fig-3}; the incomplete skyrmion looks like a vortex \cite{Shinjo-etal-science2000,Okuno-etal-JMMM2002}. Nevertheless, when the diameter $d$ increases further, an isolated skyrmion, denoted by "isolated", emerges in the positive region of $f$  (Fig. \ref{fig-5}(e)), in which the polarity between the center and periphery changes more than $\pi$ in contrast to the case of incomplete skyrmion. When $f$ is further increased to $f\!=\!2.56$, incomplete skyrmions appear even on the  $N\!=\!11932$ lattice (Fig. \ref{fig-5}(f)).

The emergence of incomplete skyrmions is expected to be owing to the strong MEC effect for a relatively large $f(>0)$, where a strain configuration ($\Leftrightarrow\! \vec{\tau}\!\propto\! \vec{e}^{\;r}\!=\! \vec{r}/r$) shares the center with the skyrmion for the rotational symmetry independent of $N$. This rotationally symmetric configuration of  $\vec{\tau}$  enhances the surface effect of GC. The $\vec{\tau}$ direction is almost perpendicular to the bond directions, and the FMI and DMI are expected to be very small on the side surface owing to the FG modeling prescription.  Therefore, there is no chance for incomplete skyrmions to appear independently of the primary incomplete skyrmions on the nanodots, which emerge as a ground state without strain under zero magnetic field \cite{Beg-etal-SRep2015,Beg-etal-PRB2017}. This problem will be further discussed below.

% subsection
\subsection{Order parameters and energies \label{order_param}}
% f-6
\begin{figure}[h]
	\centering{}\includegraphics[width=11.5cm,clip]{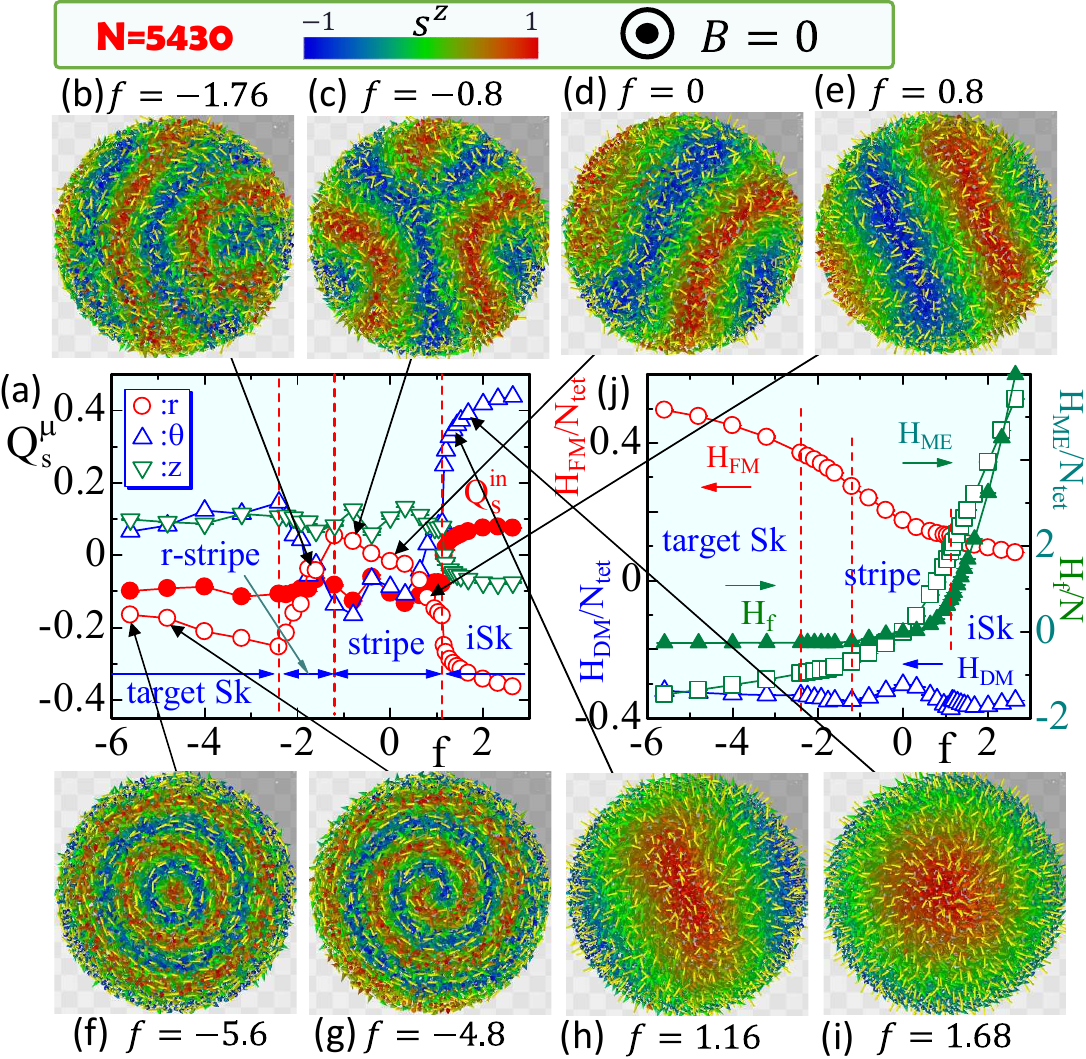}
	\caption{
	(a) Nonpolar order parameters  $Q_{s}^\mu (\mu\!=\!r,\theta,z)$ and $Q_{s}^{\rm in}(=\!Q_{s}^{r}\!+\!Q_{s}^{\theta})$  vs. $f$, where the indices represent the directions $\mu$ and spin $s$, and "in" denotes "in-plane".   (b)--(i) Snapshots for $-5.6\leq f\leq 2.64$ obtained on the $N\!=\!5430$ lattice under $B\!=\!0$, and (j) FMI, DMI, MEC, and stress-strain energies in Eq. (\ref{Hamiltonians}) vs. $f$.  We find three different phases; target skyrmion (target Sk) and stripe, and incomplete skyrmion (iSk),  and the stripe phase can be divided into two phases; rounded stripe ("r-stripe") and stripe. We find target skyrmions in (f) and (g), elongated incomplete skyrmion in (h) and rotationally symmetric incomplete skyrmion in (i), where ${s}^z$ at the periphery is $s_z\!>\!-1$. Note that $-\alpha H_{\rm ME}$ and $-H_f$ are included in the total Hamiltonian of Eq. (\ref{total-Hamiltonian}).	\label{fig-6} }
\end{figure}
In Fig. \ref{fig-6}(a), we plot nonpolar order parameters of spins defined by
\begin{eqnarray}
	\label{nonpolar_oderprm}
	\begin{split}
		 Q_{s}^\mu=&\frac{3}{2}\left(\langle({\bf s}\cdot\vec{e}_i^{\;\mu})^2\rangle-\frac{1}{3}\right),\\
		& (-0.5\leq Q_s^\mu\leq 1 ), \quad (\mu=r,\theta,z), \\
		 Q_{s}^{\rm in}=&Q_{s}^{r}+Q_{s}^{\theta}
	\end{split}
\end{eqnarray}
 where $\vec{e}_i^{\;r}$ and $\vec{e}_i^{\;\theta}$ denote unit vectors along the radial and tangential directions at vertex $i$, respectively   (Appendix \ref{App-C}). Order parameters of this type are always used for liquid crystal molecules, which are nonpolar  \cite{Andrienko-JML2018}. 
 The reason why nonpolar order parameters are calculated for ${\bf s}$ is that $Q_{s}^\mu$ are useful to find the phase boundaries as shown below.

The parameters $Q_{s}^{r}$ (\textcolor{red}{$\circ$}), $Q_{s}^{\theta}$ (\textcolor{blue}{$\triangle$}) and $Q_{s}^{\rm in}$ (\textcolor{red}{$\bullet$}) discontinuously change at $f\!\simeq\! 1$ indicating that the incomplete skyrmion phase is discontinuously separated from the stripe phase.  The jump of $Q_s^{\rm in}$ at the phase boundary is owing to those in $Q_s^{r, \theta}$. The jump in $Q_s^{\theta}$ implies an anisotropic alignment of ${\bf s}_i$ in the $\pm {\vec e}^{\;\theta}$ directions, and this in-plane magnetic anisotropy as a response to $f(>\!0)$ is a non-trivial effect of FMI, DMI, and MEC implemented in $H_{\rm FM,DM,ME}$. We find an intermediate phase denoted by "r-stripe" (rounded stripe) between the target skyrmion and stripe phases has continuous but cusp-like changes in the variation of $Q_{s}^{r}$ (\textcolor{red}{$\circ$}) and $Q_{s}^{\theta}$ (\textcolor{blue}{$\triangle$}) vs. $f$. The incomplete skyrmion and stripe phases are also characterized by $Q_s^{\rm in}$ (\textcolor{red}{$\bullet$}), although the difference in $Q_s^{\rm in}$ between the two phases is not very large.  Note that behaviors of $Q_s^{\mu}$ depend on the lattice size $N$; the transition point between the incomplete skyrmion and stripe phases depends on $N$. Here, we only describe the results obtained on the $N\!=\!5430$ lattice, which is not sufficiently large. However,  this lattice size is reasonable for studying the surface effect.

Here, we discuss the differences in $Q_{s}^r$ between the target skyrmion, r-stripe,  stripe, and incomplete skyrmion phases. Note that $Q_{s}^r\!\to\!-0.5$ if all ${\bf s}$ satisfy ${\bf s}\!\perp\! \vec{e}^{\;r}$, $Q_{s}^r\!\to\!0$ for random ${\bf s}$ with respect to the direction $\vec{e}^{\;r}$, and  $Q_{s}^r\!\to\!1$ for ${\bf s}\!\parallel\! \vec{e}^{\; r}$. These properties of $Q_{s}^\mu$ describe  directional- or anisotropic-property of magnetization independent of $\pm$-direction. In the target skyrmion phase (Figs. \ref{fig-6}(f),(g)), $Q_{s}^r$ remains monotonically decreasing  as $f$ increases but almost constant; $Q_{s}^r\!\simeq\!-0.2$, indicating that  ${\bf s}\!\perp\! \vec{e}^{\;r}$ for many ${\bf s}$. In the r-stripe phase (Fig. \ref{fig-6}(b)), $Q_{s}^r$ starts to increase  implying that the condition ${\bf s}\!\perp\! \vec{e}^{\;r}$ is not always satisfied as $f$ increases. This increase in $Q_{s}^r$ terminates and decreases again as $f$ moves to the right in the stripe phase (Figs. \ref{fig-6}(c),(d) and (e)), and it discontinuously decreases at $f\!\simeq\!1$ illuminating the phase boundaries. 

In the incomplete skyrmion phase  (Figs. \ref{fig-6}(h),(i)), we find $Q_{s}^r\!\to\!-0.5$ for large $f(>\!0)$ corresponding to ${\bf s}\!\perp\! \vec{e}^{\;r}$. This  ${\bf s}\!\perp\! \vec{e}^{\;r}$ implies that ${\bf s}$ aligns along the $z$-direction or $\theta$-direction and is reasonably understood from the incomplete skyrmion configuration. The alignment in the  $\theta$-direction can be confirmed in $Q_{s}^\theta$ in Fig. \ref{fig-6}(a), and the alignment along the  $z$-direction will be shown below. Thus, we find a strain-induced  magnetic anisotropy along the $\theta$-direction and $z$-direction in the incomplete skyrmion phase for increasing tensile stress.

The sign of $s^z$ at the periphery of the skyrmions that emerge under tensile stress is not always completely opposite to that at the center, as shown in Figs. \ref{fig-6}(h),(i). Therefore, these skyrmions resemble vortex configurations, as mentioned above \cite{Shinjo-etal-science2000,Okuno-etal-JMMM2002}. However, the opposite sign of $s^z$ at the periphery is apparent when $f$ is small; that is, $f\!=\!1.68$ in Fig. \ref{fig-6}(i). Therefore, the configurations shown in Figs. \ref{fig-6}(h),(i) are considered to be incomplete skyrmions.

 FMI, DMI and MEC energies per tetrahedron plotted in Fig. \ref{fig-6}(j), where $H_{\rm FM}$, $H_{\rm DM}$, $H_{\rm ME}$ and $H_f$  are given in Eq. (\ref{Hamiltonians}) and 
$N_{\rm tet}$ is the total number of tetrahedrons.  
We find that $H_{\rm FM}/N_{\rm tet}$ ($H_{\rm DM}/N_{\rm tet}$) in the incomplete skyrmion phase is smaller (negatively larger) than that in the other phases, implying that the incomplete skyrmion phase is energetically stable for  the FMI and DMI, although $H_{\rm DM}/N_{\rm tet}$ discontinuously increases when $f$ increases at $f\!\simeq\! 1$. $H_{\rm ME}/N_{\rm tet}$ and $H_f/N$ increase almost linearly with increasing $f$ for $f>0$ as expected from the definitions of $H_{\rm ME}$ and $H_{f}$ in Eq. (\ref{Hamiltonians}). These results also imply that $H_{\rm ME}$ and $H_{f}$ are small or negatively large in the incomplete skyrmion phase because $-\alpha H_{\rm ME}$ and $-H_{f}$ are included in the total Hamiltonian of Eq. (\ref{total-Hamiltonian}). The shape of $H_{f}/N_{\rm tet}$ is almost constant zero for $f\!<\!0$ implying that $(\vec{\tau}_i\cdot\vec{f})^2\!\to\!0$ for negatively large $f$. This result is consistent with the observation that  $\vec{\tau}_i\perp\vec{e}^{\; r}$ in the target skyrmion phase (Figs. \ref{fig-4}(a) and  \ref{fig-5}(a)-(c)).

\begin{figure}[h]
	\centering{}\includegraphics[width=10.5cm,clip]{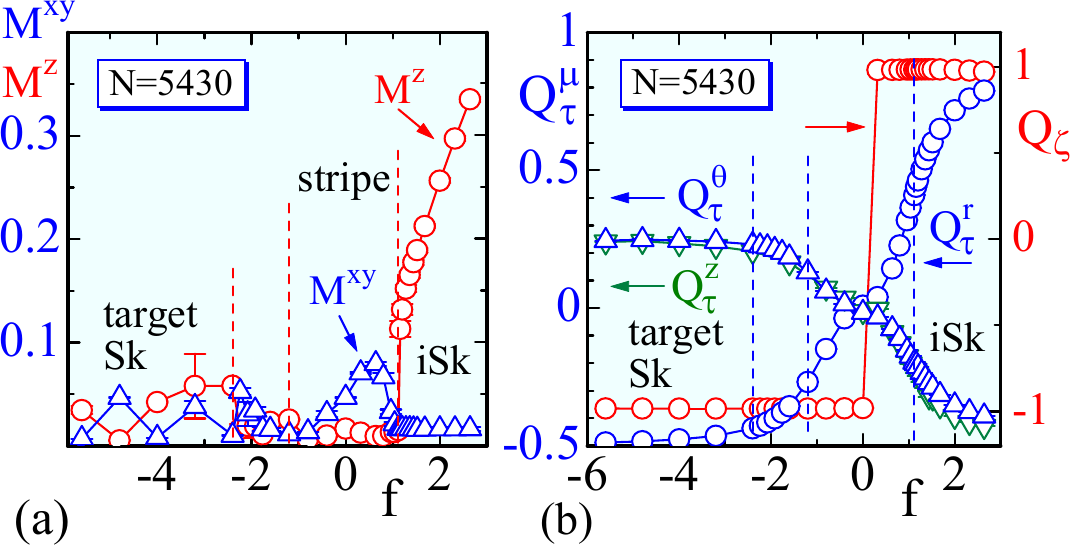}
	\caption{
		(a) Polar order parameter  $M^z$ and $M^{xy}$ vs. $f$, and (b) order parameters $Q_\zeta$ and $Q_{\tau}^{\,\mu} (\mu=r,\theta,z)$ vs. $f$. 
		\label{fig-7} }
\end{figure}

To see effects of strains on magnetization, we plot the absolute of polar order parameters
\begin{eqnarray}
	\label{polar_oderprm}
	M^z=\frac{1}{N}|\sum_i s_i^z|,  \quad M^{xy}=\frac{1}{N}\sqrt{(\sum_i s_i^x)^2+(\sum_i s_i^y)^2}
\end{eqnarray}
in Fig. \ref{fig-7}(a), where the dashed vertical lines represent the phase boundaries plotted in Fig. \ref{fig-6}(a). In the incomplete skyrmion phase, we find that $M^z$ increases from $M^z \!\simeq\!0$  to $M^z\!>\!0$ as $f$ increases. The small increment of $M^z$ with increasing $f$ is considered a signal of "the strain-induced alignment of ${\bf s}$ along the $\pm z$ direction".  This result is consistent with "the strain-mediated increase of in-plane magnetic anisotropy" observed under increasing compression in Ref. \cite{YWang-etal-NatCom2020}, because we can read the plotted $M^z$ such that $M^z$ decreases with decreasing $f$ in the incomplete skyrmion phase.  The result $M^{xy}\!\simeq\!0$ does not imply that the tensile stress $f$ decreases the in-plane magnetization  but is simply owing to the cancellation in $\pm s_i^x$ and  $\pm s_i^y$. This is the reason why the non-polar order parameters $Q_s^\mu$ plotted in Fig. \ref{fig-6}(a) are useful to observe the $\pm$-direction independent alignment. The plotted data $M^z\!\to\!0$ in the stripe and target skyrmion phases are also reasonable because ${\bf s}$ that align along the $\pm z$ direction cancel each other.
% f-7

An order parameter $Q_\zeta$ for $\zeta$ defined by 
\begin{eqnarray}
	Q_\zeta=(1/N_B)\sum_a \zeta_a, 
\end{eqnarray}
is plotted in Fig. \ref{fig-7}(b). The value $Q_\zeta\!=\!1$ ($Q_\zeta\!=\!-1$) corresponds to the cosine (sine) type $v_{ij}$ in Eq. (\ref{Finsler-unit-length-ME}). An abrupt change in $Q_\zeta$ at $f\!=\!0$ does not always indicate an internal phase transition caused by external mechanical stress $f$, because no abrupt change is observed in energies including $H_\zeta$. However, this change in  $Q_\zeta$ causes a discontinuity in the effective coupling constant $\Omega$ as shown below. Notably, we verified that no target skyrmion configuration would emerge if $Q_\zeta\!=\!1 (\Leftrightarrow\! v_{ij}\!=\!|\cos\theta_{ij}|\!+\!v_0)$ for $f\!<\!0$ by manually fixing $\zeta$ to be $\zeta\!=\!1$. Therefore, this transition in $Q_\zeta$  (Fig \ref{fig-7}(b)) is expected to be crucial for the morphological change on nanodots under radial stress. From the direction-dependent nonpolar order parameters $Q_{\tau}^\mu\!=\!\frac{3}{2}\left(\langle(\vec{\tau}\cdot\vec{e}_i^{\;\mu})^2\rangle-\frac{1}{3}\right), \; (\mu=r,\theta,z)$ in Fig. \ref{fig-7}(b), we find that the strain $\vec{\tau}$ is at random ($\Leftrightarrow\! Q_{\tau}^\mu\to\!0, (\mu\!=\!r,\theta,z)$) for $f\!\to\!0$ and aligns such that  $\tau^r\!\to\!0$ ($\Leftrightarrow\! Q_{\tau}^r\to\!-0.5$) [($\tau^{\theta,z}\!\to\!0$  ($\Leftrightarrow\! Q_{\tau}^{\theta,z}\to\!-0.5$)]  in the target [incomplete] skyrmion phase, as expected.

% subsection
\subsection{Direction-dependent interaction coefficients with effective Hamiltonian and topological charge \label{direction_dep_coeff}}
% f-8
\begin{figure}[h]
	\centering{}\includegraphics[width=13.5cm,clip]{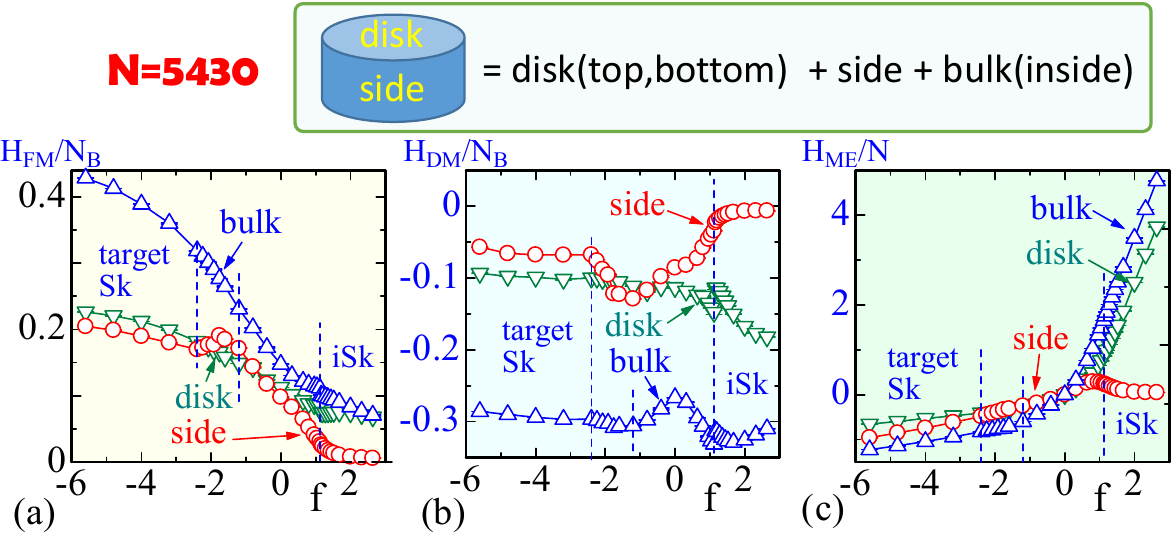}
	\caption{
		 Position-dependent energies. Bulk and side energies per bond for (a) FMI, (b) DMI, and (c) MEC. The coefficients $(\lambda,D,\alpha)\!=\!(1,0.7,0.3)$ are not included in these energies.  We find a surface effect that ${H}_{\rm FM}({\rm side})/N_B\!\to\!0$, ${H}_{\rm DM}({\rm side})/N_B\!\to\!0$	and ${H}_{\rm ME}({\rm side})/N_B\!\to\!0$ in the incomplete skyrmion (iSk) phase. We also find a competing nature of FMI and DMI in the iSk phase, such that ${H}_{\rm FM}({\rm bulk})/N_B$ decreases, whereas ${H}_{\rm DM}({\rm bulk})/N_B$ increases with increasing $f$. Note that $-\alpha H_{\rm ME}$ is included in the total Hamiltonian of Eq. (\ref{total-Hamiltonian}).   \label{fig-8} }
\end{figure}
First, we plot the FMI and DMI Hamiltonians on the side surface and inside or bulk to show the surface effect. From the identity $\sum_{\Delta} \sum_{ij(\Delta)}\!=\!\sum_{ij}\sum_{\Delta(ij)}$ for the summation convention, energies per tetrahedron can also be written as energies per bond such that 
\begin{eqnarray}
	\label{energies-per-bond}
	\begin{split}
		&	H_{\rm FM}/N_B=\frac{1}{N_B}\dot{\sum_{ij}} \sum_{\Delta(ij)}\Gamma_{ij}(\vec{\tau})\left(1-{\bf s}_i\cdot{\bf s}_j\right),\\
		&	H_{\rm DM}/N_B=\frac{1}{N_B}\dot{\sum_{ij}} \sum_{\Delta(ij)}\Gamma_{ij}(\vec{\tau})\vec{e}_{ij}\cdot {\bf s}_i\times{\bf s}_j,\\
		&  	H_{\rm ME}/N_B=\frac{f}{N_B}\dot{\sum_{ij}} \sum_{\Delta(ij)}\Omega_{ij}(\vec{\tau},\zeta)\left({\bf s}_i\cdot{\bf s}_j\right)^2, 	 	
	\end{split}
\end{eqnarray}
where $\dot{\sum}_{ij}$ denotes that the sum over bond $ij$ is limited to a part of the lattice, and $N_B(=\!\dot{\sum}_{ij}1)$ is the total number of bonds included in that part. The expression of energy per bond is more convenient for observing the difference in energies between the surface and bulk, because it is apparent whether a bond $ij$ is on the surface or not. Note that, when $H_{\rm DM}/N_B$ is limited to the side surface, $N_B$ is the total number of bonds on the side surface. The circular edges shared by the side surface and disks are one-dimensional objects; however, the edges are included on the side surface, for simplicity. 

The data obtained on the $N\!=\!5430$ lattice are plotted in  Figs. \ref{fig-8}(a)--(c).  We find a large difference between ${H}_{\rm DM}({\rm side})/N_B$ and ${H}_{\rm DM}({\rm bulk})/N_B$. One reason for this difference is that (i) there exists a sum over tetrahedrons $\dot{\sum}_{\Delta(ij)}$ inside the sum over bonds $\dot{\sum}_{ij}$ such that $\dot{\sum}_{ij} \sum_{\Delta(ij)}$ in $H_{\rm DM}$, and this ${\sum}_{\Delta(ij)}$ depends on whether bond $ij$ is on the surface or inside; the large difference is not obtained if we use the sum over bonds  $H_{\rm DM}\!=\!\dot{\sum}_{ij}\vec{e}_{ij}\cdot {\bf s}_i\times{\bf s}_j$. The other reason is a dynamical one that (ii) the intensive part of energy $\Gamma_{ij}(\vec{\tau})$ included in  $H_{\rm DM}$ depends on position and direction. This second part (ii) is a main topic in this paper. 

% f-9
\begin{figure}[h]
	\centering{}\includegraphics[width=13.5cm,clip]{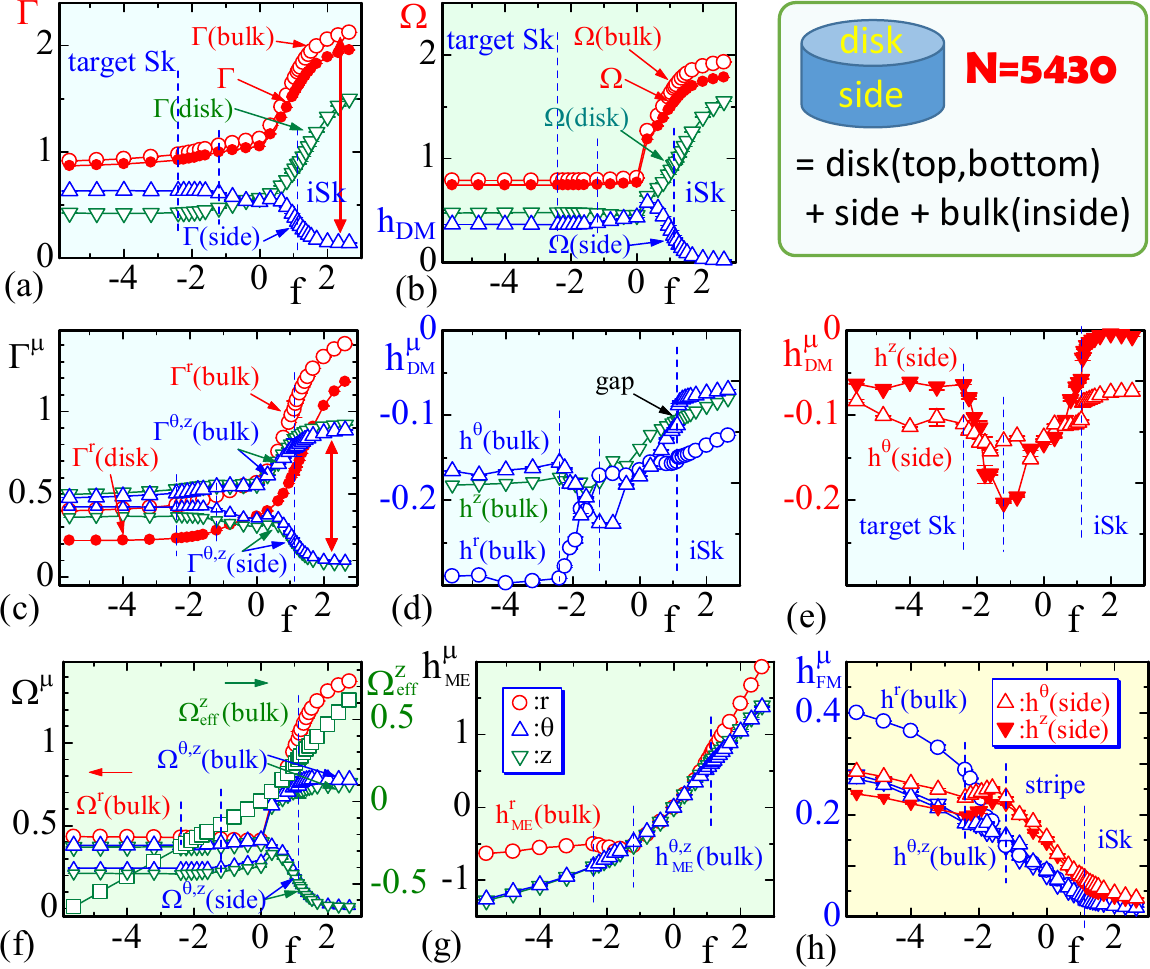}
	\caption{
		(a) Position-dependent $\Gamma ({\rm bulk, side, disk})$ in Eq. (\ref{effective-coupling-constant}); $\Gamma$  (\textcolor{red}{$\bullet$}) is calculated using the whole bonds. (b)  Position-dependent $\Omega ({\rm bulk, side, disk})$ for ME. (c) Direction-dependent coefficients $\Gamma^{\mu}({\rm bulk})$,  and (d) the corresponding bulk Hamiltonian  $h_{\rm DM}^{\mu}$ in Eq. (\ref{direction-dependent-effective-energies}) and (e) $h_{\rm DM}^{\mu}({\rm side})$ on the side surface.  (f) The bulk coefficient of the magneto-elastic coupling  $\Omega^{\mu}$, and (g) the corresponding bulk Hamiltonian $h_{\rm ME}^{\mu}({\rm bulk})$, (h) FMI bulk Hamiltonian $h_{\rm FM}^{\mu}({\rm bulk})$ and side Hamiltonian $h_{\rm FM}^{\mu}({\rm side})$. \label{fig-9} }
\end{figure}
To see this second reason (ii), we plot the effective DMI (and FMI) coupling constant defined by
\begin{eqnarray}
	\label{mean-DMI-coefficient}
	\Gamma=\frac{1}{N_B}\dot{\sum}_{ij}\sum_{\Delta(ij)}\Gamma_{ij}(\vec{\tau})
\end{eqnarray}
and $\Omega$ defined by the same expression in Figs. \ref{fig-9}(a),(b). 
  In this lattice average, the direction dependence of $\Gamma_{ij}(\vec{\tau})$ is integrated out or summed over. To extract the direction dependence of $\Gamma_{ij}(\vec{\tau})$, we define the direction-dependent coefficients 
\begin{eqnarray}
	\label{effective-coupling-constant}
	\begin{split}
		&	\Gamma^\mu=\frac{1}{N_B}\dot{\sum}_{ij} \sum_{\Delta(ij)}|\Gamma_{ij}^{\mu}(\vec{\tau})|, \;(\mu=r,\theta,z)
	\end{split}
\end{eqnarray}
which are plotted in Fig.  \ref{fig-9}(c), where $\Gamma_{ij}^{\mu}(\vec{\tau})$ is given by Eq. (\ref{Gamma-decomposition}). The corresponding direction-dependent energies $h_{\rm  DM}^{\mu}$ plotted in Fig.  \ref{fig-9}(d) are defined by
\begin{eqnarray}
	\label{direction-dependent-effective-energies}
	\begin{split}
		&	h_{\rm DM}^{\mu}=H_{\rm DM}^{\mu}/N_B, \\
		&	H_{\rm DM}^{\mu}=\frac{\dot{\sum}_{ij} \sum_{\Delta(ij)}\Gamma_{ij}(\vec{\tau})(\vec{e}_{ij}\cdot\vec{e}^{\;\mu})^2\vec{e}_{ij}\cdot {\bf s}_i\times{\bf s}_j}{\frac{1}{N_B}\dot{\sum}_{ij} \sum_{\Delta(ij)}|\Gamma_{ij}^{\mu}(\vec{\tau})|}, \;(\mu=r,\theta,z)
	\end{split}
\end{eqnarray}
where $\sum_{\Delta}\sum_{ij(\Delta)}$ is replaced by $\dot{\sum}_{ij} \sum_{\Delta(ij)}$ in $H_{\rm DM}^{\mu}$ of Eq. (\ref{HDMI-decomposition}) and $N_B$ is given by $N_B\!=\!\dot{\sum}_{ij}1$  as in Eq. (\ref{energies-per-bond}).

The reason for $\Gamma^\mu\!\simeq\!0.5$ at $f\!\to\!0$ in Fig. \ref{fig-9}(c) is that the normalization factor $\hat{\Gamma}$ of $\Gamma_{ij}$ in Eq. (\ref{Gamma-X}) is defined such that $\Gamma_{ij}\!\to\!1$  for isotropic configuration of $\vec{\tau}$ and that the lattices are constructed isotropic such that the mean value of $|\vec{e}_{ij}\cdot \vec{e}^{\;\mu}|$ is $|\vec{e}_{ij}\cdot \vec{e}^{\;\mu}|\!\simeq\!0.5$. The large difference between  $\Gamma^{\rm r}({\rm bulk})$ and $\Gamma^{\theta,z}({\rm bulk})$ in the incomplete skyrmion phase shown in Fig.  \ref{fig-9}(c) indicates a strong helical order along the $\vec{e}^{\;r}$ direction than along the $\vec{e}^{\;\theta,z}$ directions.  Discontinuities of $\Omega$ and $\Omega({\rm bulk})$ at $f\!\to\!0$ observed in Fig. \ref{fig-9}(b) are owing to the difference between $v_{ij}(\zeta_{ij}\!=\!1)$ and $v_{ij}(\zeta_{ij}\!=\!-1)$ in Eq. (\ref{Finsler-unit-length-ME}) corresponding to the discontinuous change in $Q_\zeta$ plotted in Fig. \ref{fig-7}(b). 

The surface DMI ($=\!\Gamma(\rm side)$  and $\Gamma^{\theta,z}({\rm side})$) plotted in the lower part of  Figs.  \ref{fig-9}(a) and  \ref{fig-9}(c) are remarkably smaller than the bulk DMI ($=\!\Gamma({\rm bulk})$ and $=\!\Gamma^{\theta,z}({\rm bulk})$) indicated by (\textcolor{red}{$\updownarrow$}) in the incomplete skyrmion phase; the $r$ component is not defined on the side surface because $\vec{e}_{ij} \perp\vec{e}^{\; r}$ is expected on the side surface. Note also that the direction dependence or anisotropy  and the position dependence of $\Gamma^\mu$ are apparent only for $f\!>\!0$, where $\vec{\tau}\!\propto\! \vec{e}^{\; r}$ (Figs. \ref{fig-4}(c), \ref{fig-5}(f)). 
This dynamical enhancement of the surface effects is considered to be an origin of the GC effect on the skyrmions in nanodots and compatible with the assumption of zero surface DMI in a confinement model of Ref. \cite{Diguet-etal-JMMM2023}. By contrast, the direction and position dependence of  $\Gamma^\mu$ disappears for negative large $f(<0)$, where $\vec{\tau}$ is spiral along the $z$ direction (Figs. \ref{fig-4}(a), \ref{fig-5}(a)--(c)).

 No apparent difference is observed between the surface DMI ($=\!\Gamma^{\rm r}({\rm disk})$) on the upper/lower disk  and the bulk DMI ($=\!\Gamma^{\rm r}({\rm bulk})$). The other component $\Gamma^{\rm \theta}({\rm disk})$ that is not plotted is comparable to the bulk component $\Gamma^{\rm \theta}({\rm bulk})$. Note that $\Gamma^\mu, (\mu=r, \theta,z)$ are considered the $\mu$ components  of a DMI vector \cite{Fert-etal-NatReview2017,Zhang-JPCM2020,Tokura-Kanazawa-ACS2020,Gobel-etal-PhysRep2021,Wang-etal-JMMM2022} (Appendix \ref{App-D}). It is also noteworthy that $\Gamma^{\mu}({\rm bulk}), (\mu\!=\!r,\theta,z)$ increase with increasing $f$ in the whole range of $f$ and this behavior is qualitatively consistent with the reported result that $D_{\rm ave}$ decreases with increasing strain in Ref.  \cite{YWang-etal-NatCom2020}, where $-D_{\rm ave}$ corresponds to  $\Gamma$ plotted in Fig. \ref{fig-9}(a) with the symbol (\textcolor{red}{$\bullet$}).

The direction dependence of $h_{\rm DM}^{\mu}({\rm bulk})$ (Fig. \ref{fig-9}(d)) is apparent in the target and incomplete skyrmion phases. The direction-dependent energies $h_{\rm DM}^\mu({\rm side})$ plotted in Fig. \ref{fig-9}(e) are comparable with the bulk ones $h_{\rm DM}^\mu({\rm bulk})$ in \ref{fig-9}(d). 
This result implies that the reason for a large difference between $h_{\rm DM}({\rm side})$ and $h_{\rm DM}({\rm bulk})$ in Fig. \ref{fig-8}(b) is mainly due to the difference between  $\Gamma^{\mu}({\rm side})$ and  $\Gamma^{\mu}({\rm bulk})$, because the DMI energies plotted in Fig. \ref{fig-8}(b) can also be expressed by using the direction-dependent quantities in Eq. (\ref{HDM-decomp-rtz}) such that 
\begin{eqnarray}
	\label{HDM-decomp-rtz-text}
	&	 H_{\rm DM}=\Gamma^{r} H_{\rm DM}^{r}+\Gamma^{\theta} H_{\rm DM}^{\theta}+\Gamma^{z} H_{\rm DM}^{z}.
\end{eqnarray} 
Thus, we should emphasize that the reason for $|H_{\rm DM}({\rm side})/N_B|\!\ll\! |H_{\rm DM}({\rm bulk})/N_B|$ in Fig. \ref{fig-8}(b) is  that the intensive part of energy satisfies 
$$\Gamma^{\theta,z}({\rm side}) \ll \Gamma^{\theta,z} ({\rm bulk})$$
as shown in Fig. \ref{fig-9}(c). Note that a small discontinuity in $h_{\rm DM}^{\theta}({\rm bulk})$ at the phase boundary between the skyrmion and stripe, denoted by "gap" in  Fig. \ref{fig-9}(d), indicates a weak first-order transition.

It is also interesting to observe that the variations of $\Omega^{\mu}$ in  Fig.  \ref{fig-9}(f) are almost the same as those of  $\Gamma^{\mu}$; the definition of $\Omega_{ij}$ is considerably different from that of  $\Gamma_{ij}$ (Appendix \ref{App-B}). $\Omega^{\theta,z}({\rm side})$ are very small compared with $\Omega^{\theta,z}({\rm bulk})$ in the incomplete skyrmion phase (Fig. \ref{fig-9}(f)).  The corresponding $h_{\rm ME}^{\mu}$ plotted in Fig.  \ref{fig-9}(g) are direction-dependent only in the target and incomplete skyrmion phases similar to $h_{\rm DM}^{\mu}$. 
The symbol $\Omega_{\rm eff}^{z}$ denotes an effective coupling constant 
\begin{eqnarray}
	\Omega_{\rm eff}^{z}=\alpha f \Omega^z,
\end{eqnarray}
where the constant $\alpha$ is included. This  $\Omega_{\rm eff}^{z}$ corresponds to the magnetic anisotropy $K_{\rm ave}$ in Ref. \cite{YWang-etal-NatCom2020}, however, the $z$ component of  $H_{\rm ME}$ does not always play a role in the magnetic anisotropy in our model. Nevertheless, we find qualitative consistency between the observation that $\Omega_{\rm eff}^{z}$ increases with increasing $f$  and the behavior of $K_{\rm ave}$ versus strain  in Ref. \cite{YWang-etal-NatCom2020}. Note that $\Omega_{\rm eff}^{z}\!\to\!0$ for $f\!\to\!0$, and therefore no spontaneous magnetic anisotropy is implemented in the model, as mentioned in Section \ref{Hamiltonian}; all anisotropies are dynamically generated and can be evaluated. 

We should note that a strong anisotropy in $\Omega^{\mu}({\rm bulk}), (\mu\!=\!r,\theta,z)$, which is considered a strain-induced interaction anisotropy, appears only in the region of $f\!>\!0$ in the proposed model as confirmed from Fig. \ref{fig-9}(f).  In the region of $f\!>\!0$, we find  $\Omega^{r}({\rm bulk})>\Omega^{\theta,z}({\rm bulk})$, which indicates that the tensile stress makes the spin correlation $({\bf s}_i\cdot{\bf s}_j)^2$ along the $r$-axis to be stronger than those along the other axes. This strong and anisotropic correlation implemented in $\Omega^{r}({\rm bulk})$ of MEC, in addition to the enhanced effective FMI, competes with the enhanced DMI along the $r$-axis. Consequently, the spin correlation along the $r$ direction becomes stronger than that without MEC. This strain-enhanced competition modified by ME along the $r$-axis increases the skyrmion size and is considered the reason for the appearance of incomplete skyrmions. Note that the inequality $\Omega^{r}({\rm bulk})>\Omega^{\theta,z}({\rm bulk})$ does not imply anisotropy of ${\bf s}_i$ along the $r$ direction because $Q_s^{r}\!\to\!-0.5$ in Fig. \ref{fig-6}(a). In contrast, strain-mediated magnetic anisotropies along the $\theta,z$ directions are expected in the incomplete skyrmion phase owing to the observation that $\Gamma^{\theta,z}({\rm bulk})$ and  $\Omega^{\theta,z}({\rm bulk})$ in the incomplete skyrmion phase are larger than those in the stripe and target skyrmion phases (Figs. \ref{fig-9}(c),(f)). This expectation of strain-mediated magnetic anisotropy is consistent with the increase in both $M^z$ (Fig. \ref{fig-7}(a)) and  $Q_s^\theta$ (Fig. \ref{fig-6}(a)) with increasing $f$ in the incomplete skyrmion phase. 

We also find  that $h_{\rm DM}^{\mu}({\rm bulk})$  increase as shown in Fig.  \ref{fig-9}(d), while $h_{\rm FM}^{\mu}({\rm bulk})$ decrease as shown in Fig.  \ref{fig-9}(h), as $f$ increases in the incomplete skyrmion phase. This behavior represents the competing nature of the interactions necessary for skyrmion emergence and is consistent with the observation in the full Hamiltonians $H_{\rm FM}({\rm bulk})/N_B$ and $H_{\rm DM}({\rm bulk})/N_B$  in Figs. \ref{fig-8}(a),(b).

% f-10
\begin{figure}[h]
	\centering{}\includegraphics[width=10.5cm,clip]{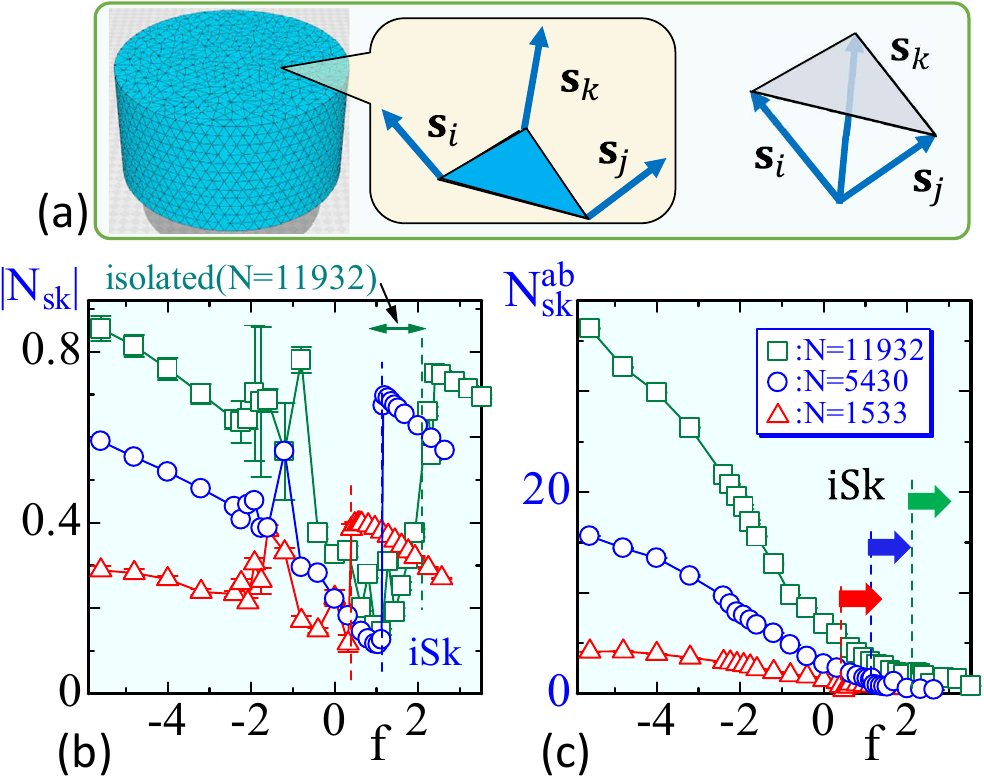}
	\caption{
		(a) A triangle $ijk$ and the spins ${\bf s}_i$, ${\bf s}_j$ and ${\bf s}_k$ for the calculation of the topological charge $N_{\rm sk}$ on the upper disk. (b) $|N_{\rm sk}|$ vs. $f$ and (c) absolute topological charge $N_{\rm sk}^{\rm ab}$ vs. $f$. The dashed vertical lines are discontinuous phase boundaries between incomplete skyrmion and stripe phases ($N\!=\!1533$, $N\!=\!5430$), and a continuous phase boundary between incomplete skyrmion and isolated skyrymion phases ($N\!=\!11932$). The dashed vertical lines represent the phase boundaries depending on $N$.   \label{fig-10} }
\end{figure}
The topological charge
$	N_{\rm sk}\!=\!\frac{1}{4\pi}\int d^2x\, {\bf s}\!\cdot\! \frac{\partial {\bf s}}{\partial x^1}\!\times\!\frac{\partial {\bf s}}{\partial x^2}$ 
is calculated using the discrete expression
\begin{eqnarray}
	\label{top-charge}
	N_{\rm sk}=\frac{1}{4\pi}\sum_{\it \Delta(ijk)} {\bf s}_i\cdot \left({\bf s}_j-{\bf s}_i\right)\times\left({\bf s}_k-{\bf s}_i\right)
\end{eqnarray}
on the upper disk of the lattice, where $\sum_{\it \Delta(ijk)}$ denotes the sum of triangles with vertices $i$, $j$ and $k$ (Fig. \ref{fig-10}(a)).  We observe a topological phase transition between the skyrmion and stripe phases (Fig. \ref{fig-10}(b)) at least for $N\!\leq\!5430$, although $|N_{\rm sk}|$ is smaller than $|N_{\rm sk}|\!=\!1$  in the skyrmion phase owing to the nature of incomplete skyrmions that the polarity change between the center and the periphery is less than $\pi$. Another reason for $|N_{\rm sk}|\!<\!1$ is the numerical error originating from discretization.  The order of the transition is considered first-order if we employ the definition of the first-order transition such that there exists a physical quantity that is discontinuously changing. Moreover,  $|N_{\rm sk}|$ in the skyrmion phase for the $N\!\geq\!5430$ lattices is larger than  $|N_{\rm sk}|(=\!0.5)$ of the vortex, implying that the skyrmion configurations under tensile stress are different from those of vortices, as discussed in Section \ref{order_param}. Note that an isolated skyrmion phase appears on the $N\!=\!11932$ lattice between the incomplete skyrmion and stripe phases in the region of $f$  indicated by (\textcolor{teal}{$\leftrightarrow$}) in Fig. \ref{fig-10}(a). However, surface effects are expected to be weak due to the increasing bulk volume on large nanodots, and for this reason, a discontinuous transition is observed only on smaller nanodots.

The absolute of topological charge defined by
\begin{eqnarray}
	\label{top-charge-ab}
	N_{\rm sk}^{\rm ab}=\frac{1}{4\pi}\sum_{\it \Delta(ijk)} |{\bf s}_i\cdot \left({\bf s}_j-{\bf s}_i\right)\times\left({\bf s}_k-{\bf s}_i\right)|
\end{eqnarray}
is convenient to see the total topological excitation (Fig.  \ref{fig-10}(c)). This quantity is called absolute skyrmion number in Ref. \cite{Beg-etal-SRep2015,Beg-etal-PRB2017}. We find that $N_{\rm sk}^{\rm ab}$ increases with increasing $N$ and smoothly varies with respect to $f$.

% Section
\section{Concluding remarks \label{conclusion}}
In this paper, we numerically study the skyrmion stability in nanodots with and without a magnetic field using the Finsler geometry (FG) modeling technique and find that a single incomplete and/or isolated skyrmion is stabilized at the center of the nanodots under radial tension. Metropolis Monte Carlo technique is used for the simulations on 3D tetrahedral lattices of cylindrical shape.  Our Hamiltonian is defined with the variables of electric spins and strain direction, and an Ising-like variable to dynamically treat the anisotropy of the magneto-elastic coupling (MEC). In such FG modeling, position- and direction-dependent coupling constants emerge dynamically in response to the the tensile stress.

As a GC effect without stress, we confirm that the minimal external magnetic field for skyrmions decreases when the nanodot diameter decreases. This observation is consistent with  previously reported experimental data by Y. Wang et al. \cite{YWang-etal-NatCom2020} for materials with a spontaneous in-plane magnetic anisotropy. When radial stresses are applied under zero magnetic field, we confirm three different phases: incomplete (or isolated) skyrmions, stripes, and target skyrmions, by varying the stress from positive (tension) to negative (compression). In the incomplete skyrmion phase, we find that the effective DMI coefficient increases with increasing tensile stress, and there is a large difference in the effective coupling constants, including FMI and MEC, between the surface and bulk in the nanodots. This modification of the coupling constants is considered a strain-enhanced surface effect. We also find a strain-induced $z$-direction alignment of spin configurations in the polar-order parameter. These observations are consistent with the experimental results reported in Ref. \cite{YWang-etal-NatCom2020} that the DMI coefficient and spontaneous in-plane magnetic anisotropy are modified by strains. Moreover, we observe that the incomplete skyrmions close to the stripe phase change to isolated skyrmions when the disk diameter increases.

Thus, we find that a stable incomplete (or isolated) skyrmion emerges at the center of the nanodot owing to the dynamical enhancement of the surface effect by radially applied tensile stresses. In summary, the origins of skyrmion stability in nanodots without a magnetic field are (i) dynamically enhanced or strain-enhanced surface effects for position-dependent interactions and (ii) dynamically generated or strain-induced interaction anisotropies of the FMI, DMI, and MEC by radial stresses.

%----------------------------------------------------------
\section{Data Availability Statement}
%----------------------------------------------------------
The data that support the findings of this study are available from the corresponding author upon reasonable request.

\begin{acknowledgments}
This work is supported in part by Collaborative Research Project J23Ly07 of the Institute of Fluid Science (IFS), Tohoku University. Numerical simulations were performed on the supercomputer system AFI-NITY at the Advanced Fluid Information Research Center, IFS, Tohoku University.
\end{acknowledgments}

\appendix

% App A
\section{Lattice construction \label{App-A}}
% f-A-1 
\begin{figure}[h]
	\centering{}\includegraphics[width=10.5cm,clip]{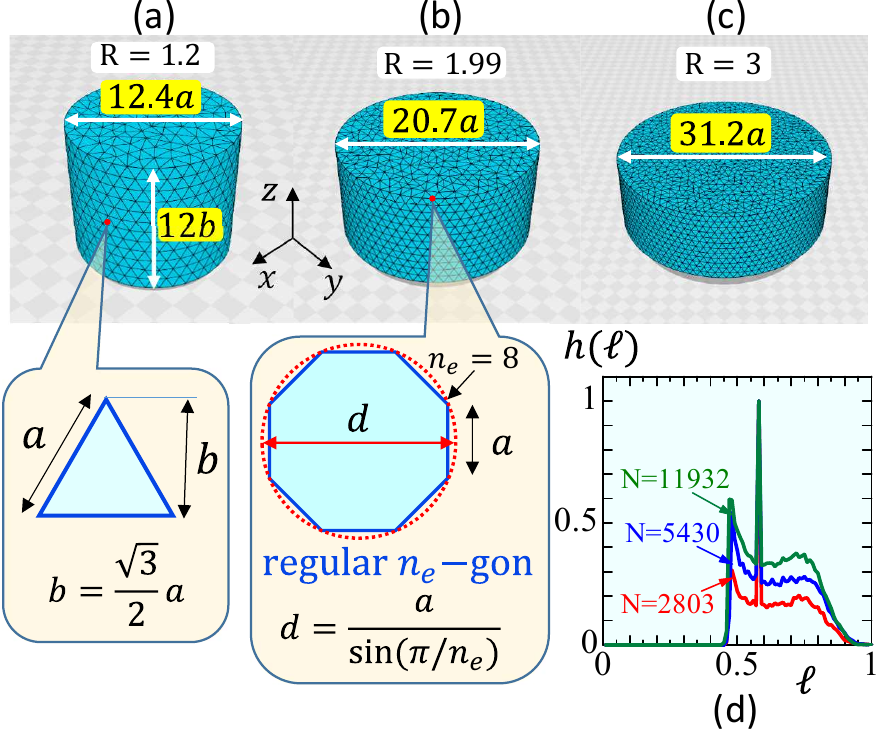}
	\caption{
		Cylindrical lattices of size (a) $N\!=\!2803$, (b) $N\!=\!5430$ and  (c) $N\!=\!11932$ composed of tetrahedrons.  The diameter $d$ and the height $h$ are written in the plots; $h$ is fixed to $h\!=\!12b\!=\!6\sqrt{3}a$, where $a$ is the edge length of regular triangle on the side cylinder and $b$ is the height. The diameter $d\!=\!a/\sin(\pi/n_e)$ is that of circumscribed circle of regular $n_e$-gon with $n_e$ given in Table \ref{table-1}, because the upper and lower disks are identified with the regular $n_e$-gon. (d) Distribution $h(\ell)$  of bond length $\ell$, where both $h(\ell)$ and $\ell$ are normalized. The peak at  $\ell\!\simeq\! 0.6$ corresponds to the bond of the cylindrical side surface. The height shares the symbol $h$ with the distribution $h(\ell)$, however, no confusion is expected.	\label{fig-A-1} }
\end{figure}
%++++++++++++++++++++++++++++++++++
\begin{table}[htb]
	\caption{Lattices are characterized by the numbers $N, N_B,N_T,N_{\rm tet}$, which are the total number of vertices, bonds, triangles, and tetrahedrons. The Euler number    $\chi(=\!N\!-\!N_B\!+\!N_T\!-\!N_{\rm tet})$  satisfies $\chi\!=\!1$, implying that the lattices are filled with tetrahedrons. $n_e$ is the total number of edges of the upper/lower disk, which is a regular $n_e$-gon. The diameter $d$ and ratio $R(=\!d/h)$ characterize the lattice shape. The height $h$ of the cylinders is $h\!=\!12b\!=\!6\sqrt{3}a$ (Fig. \ref{fig-A-1}).  \label{table-1}} 
	\begin{center}
		\begin{tabular}{ccccccccccccccccccc}
			\hline
			\hline
		$N$ && $N_B$ && $N_T$ && $N_{\rm tet}$ && $\chi$ && $n_e$ && $d$ && $R(=d/h)$\\
			\hline
		1533  && 9970  && 16315   &&  7877    && 1    &&  33 &&   $10.5a$  && $1.02$ \\
		2083  && 13686 && 22501   && 10897    && 1    &&  39 &&	  $12.4a$  && $1.20$ \\
		5430  && 36720 && 61093   && 29802    && 1    &&  65 &&   $20.7a$  && $1.99$  \\
		8465  && 57882 && 96700   && 47282    && 1    &&  82 &&   $26.1a$  && $2.51$  \\
		11932 && 82097 && 137492  && 67326    && 1    &&  98 &&   $31.2a$  && $3.00$  \\
            \hline
			\hline
		\end{tabular}
	\end{center}
\end{table}
%++++++++++++++++++++++++++++++++++
 The height $h$ of the cylinder is fixed to $h\!=\!12b$ in the unit of regular triangle height $b(=(\sqrt{3}/2)a)$ with the edge length $a$ (Fig. \ref{fig-A-1}). The upper and lower disks are identified with a regular $n_e$-gon, where $n_e$ is the total number of vertices or edges of the disk, and the diameter $d$ is given by the circumscribed circle diameter $d\!=\!a/\sin(\pi/n_e)$. Therefore, the cylinders are also characterized by the ratio $R\!=\!d/h$. We plot three cylinders of $R$ approximately given by $R\!=\!1.20$, $R\!=\!1.99$ and $R\!=\!3.00$ in Figs. \ref{fig-A-1}(a)--(c). The cylindrical surface is composed of a regular triangle. The total number of vertices on each of the top and bottom disks, including the edge, is given by $\pi (d/2)^2/(\sqrt{3}/2)$, where $\sqrt{3}/2$ is  the area of two regular triangles in the case of  $a\!=\!1$,  and the vertices are randomly distributed as uniformly as possible. The distribution of vertices on the upper disk is the same as that on the bottom disk. The vertices inside the surface are also randomly distributed as uniformly as possible, and the total number is calculated using the volume of vertices inside the surface, in which the space is assumed to be filled with regular tetrahedrons.  Vertex volume is calculated as $(\sqrt{2}/48)\!\times\!22.8$, where $(\sqrt{2}/48$ is a quarter of the regular tetrahedron volume and $22.8$ is an approximate number of the regular tetrahedrons emanating from the inner vertices calculated by $4\pi/\Omega$ with the solid angle $\Omega$ of a vertex of the tetrahedron. The peak of the distributions $h(\ell)$ of the bond length $\ell$ plotted in \ref{fig-A-1}(d) corresponds to the bonds on the cylindrical surface.  Table \ref{table-1} lists the numbers characterizing the lattices.

The mean value of the total number of tetrahedrons $n_{ij}(=\!\sum_{\Delta(ij)}1)$ sharing bond $ij$ are shown in Table \ref{table-2}. We find that $\frac{\overline{n}(\rm{side})}{\overline{n}(\rm{bulk})}\!\simeq\! 0.51$.  This ratio is one of the origins of the surface effect on GC that the surface DMI on the side surface is smaller than the bulk DMI. The surface effect will be described in Appendix \ref{App-B}. To show that the direction $\vec{e}_{ij}$ of bond $ij$ is isotropic,  the lattice averages $(\vec{e}_{ij}\cdot \vec{e}^{\;r})^2$, $(\vec{e}_{ij}\cdot \vec{e}^{\;\theta})^2$ and $(\vec{e}_{ij}\cdot \vec{e}^{\;z})^2$ are listed. We find from $(\vec{e}_{ij}\cdot \vec{e}^{\;\mu})^2\!\simeq\!1/3, (\mu\!=\!r,\theta,z)$ that  the bond direction $\vec{e}_{ij}$ is isotropically distributed. Interaction anisotropy comes only from a dynamical reason in contrast to the surface effect.
%++++++++++++++++++++++++++++++++++
\begin{table}[htb]
	\caption{The mean value of the total number of tetrahedrons sharing bond $ij$ are calculated using $\overline{n}=\!\frac{\sum_{ij}\sum_{\Delta(ij)}1}{\sum_{ij}1}$ on upper and lower disks, side surface and inside, denoted by $\overline{n}(\rm{disk})$, $\overline{n}(\rm{side})$ and  $\overline{n}(\rm{bulk})$, respectively.  $\overline{n}$ is the mean value for all bonds $ij$. $(\vec{e}_{ij}\cdot \vec{e}^{\;\mu})^2, (\mu\!=\!r,\theta,z)$ are the lattice averages of squared $\mu$ components of bond direction $\vec{e}_{ij}$, where  $\vec{e}^{\;\mu} (\mu\!=\!r,\theta,z)$ denote the $(r,\theta,z)$ direction at the center of bond $ij$ (see Appendix \ref{App-C}).   \label{table-2}} 
	\begin{center}
		\begin{tabular}{ccccccccccccccccccc}
			\hline
			\hline
			$N$ && $\overline{n}(\rm{disk})$ && $\overline{n}(\rm{side})$ && $\overline{n}(\rm{bulk})$ && $\overline{n}$ && $(\vec{e}_{ij}\cdot \vec{e}^{\;r})^2$  && $(\vec{e}_{ij}\cdot \vec{e}^{\;\theta})^2$  && $(\vec{e}_{ij}\cdot \vec{e}^{\;z})^2$\\
			\hline
	        1533  && 2.87 && 2.64 &&  5.15   && 4.74   && 0.334  && 0.337  && 0.326   \\
	        2083  && 2.85 && 2.65 &&  5.16   && 4.78   && 0.340  && 0.336  && 0.330   \\
            5430  && 2.80 && 2.64 &&  5.17   && 4.87   && 0.333  && 0.330  && 0.338   \\
            8465  && 2.80 && 2.65 &&  5.17   && 4.90   && 0.329  && 0.329  && 0.340   \\
            11932 && 2.80 && 2.65 &&  5.17   && 4.92   && 0.330  && 0.329  && 0.342   \\
			\hline
			\hline
		\end{tabular}
	\end{center}
\end{table}
%++++++++++++++++++++++++++++++++++

\vfill\eject

% App B
\section{Discretization of direction-dependent interaction coefficient \label{App-B}}
% f-B-1
\begin{figure}[h]
	\centering{}\includegraphics[width=10.5cm,clip]{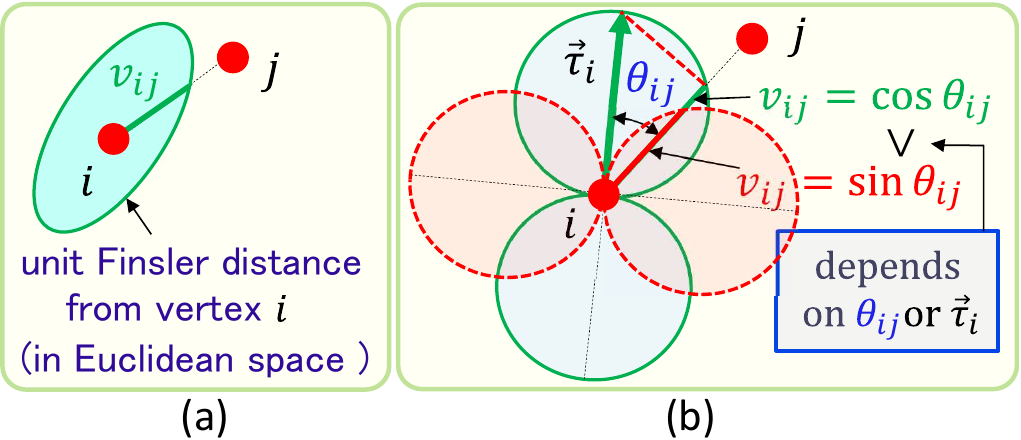}
	\caption{
		(a) Trajectory of unit Finsler length $v_{ij}$ at $i$ is not always spherical or symmetric in Euclidean space, (b) the trajectory for $v_{ij}$ assumed in this paper, where $v_0\!=\!0$ is assumed for simplicity. In (b),  $v_{ij}\!=\!\cos\theta_{ij}\!>\!v_{ij}\!=\!\sin\theta_{ij}$ for the given strain direction $\vec{\tau}_i (\in S^2/2)$. Finsler length $v_{ij}$ does not  correspond to the real length in materials but it is used to define direction-dependent interaction anisotropy between spins ${\bf s}$ and strains $\vec{\tau}$.  
		 \label{fig-B-1} }
\end{figure}
Finsler geometry framework provides interaction anisotropy between two particles via a direction-dependent Finsler length $v_{ij}$ (Fig. \ref{fig-B-1}(a)). The Finsler length $v_{ij}$ for $H_{\rm FM}, H_{\rm DM}$ is fixed to be cosine type, and $v_{ij}$ for $H_{\rm ME}$ is defined to be dynamically convertible to cosine type or sine type depending on $\zeta_{ij}$ ($\in \{-1,1\}$ on bond $ij$):
\begin{eqnarray}
	\label{Finsler-unit-length-FDMI}
&&	v_{ij}=|\cos\theta_{ij}| + v_0,\quad (v_0=0.1; \; H_{\rm FM}, H_{\rm DM}) \\
&&		v_{ij}=\left\{ \begin{array}{@{\,}ll}
	|\cos\theta_{ij}| + v_0,\;  (\zeta_{ij}=1) \\
	\sin\theta_{ij} + v_0,\;  (\zeta_{ij}=-1)
	\label{Finsler-unit-length-ME}
	\end{array} 
\right., (v_0=0.2;\;  H_{\rm ME}), 
\end{eqnarray}
where $\cos\theta_{ij}$ and $\sin\theta_{ij}$ (Fig. \ref{fig-B-1}(b)) are
\begin{eqnarray}
	\label{Finsler-coeffs}
\cos\theta_{ij}=\vec{\tau}_i\cdot\vec{e}_{ij},\quad \sin \theta_{ij}=\sqrt{1-|\vec{\tau}_i\cdot\vec{e}_{ij}|^2}\;(\geq 0),
\end{eqnarray}
and  $v_0$ is a small cutoff; the smaller $v_0$, the larger interaction anisotropy.

Using these $v_{ij}$, the Finsler metric is defined by 
\begin{eqnarray}
	\label{F-metric}
	g_{ab}=\begin{pmatrix}v_{12}^{-2} & 0 & 0\\
		0 & v_{13}^{-2} & 0 \\
		0 & 0 & v_{14}^{-2}
	\end{pmatrix},
	\quad \sqrt{g}=\sqrt{\det g_{ab}}=v_{12}^{-1}v_{13}^{-1}v_{14}^{-1}, \quad g^{ab}=(g_{ab})^{-1}.
	\label{Finsler-metric}
\end{eqnarray}
These expressions are used in the continuous Hamiltonians
\begin{eqnarray}
	\label{cont_FDM}
	\begin{split} & H_{{\rm FM}}=\frac{1}{2}\int\sqrt{g}d^{3}xg^{ab}\frac{\partial{\bf s}}{\partial x^{a}}\cdot\frac{\partial{\bf s}}{\partial x^{b}},\\
		& H_{{\rm DM}}=\int\sqrt{g}d^{3}xg^{ab}\frac{\partial{\vec{r}}}{\partial x^{a}}\cdot{\bf s}\times\frac{\partial{\bf s}}{\partial x^{b}},\\
        & H_{{\rm ME}}=\frac{1}{4}\int\sqrt{g}d^{3}x\left(g^{ab}\frac{\partial{\bf s}}{\partial x^{a}}\cdot\frac{\partial{\bf s}}{\partial x^{b}}\right)^2,		
	\end{split}
\end{eqnarray}
where the factor $f$ is removed from $H_{\rm ME}$ for simplicity. From these continuous Hamiltonians, we obtain the discrete ones in Eq. (\ref{Hamiltonians}) by replacing  the differentials with differences such that $\partial_1{\bf s}\!\to\! {\bf s}_2\!-\!{\bf s}_1$,  $\partial_2{\bf s}\!\to\! {\bf s}_3\!-\!{\bf s}_1$ and  $\partial_3{\bf s}\!\to\! {\bf s}_4\!-\!{\bf s}_1$ for a local coordinate system with the origin at the vertex $1$, and by replacing $\int\sqrt{g}d^{3}x$ with  $\sum_{\it \Delta}\sum_{ij(\Delta)}$; the sum over tetrahedrons $\sum_{\it \Delta}$ and the sum over bonds $\sum_{ij(\Delta)}$ of the tetrahedron $\Delta$, and by summing over all symmetric expressions obtained by the cyclic replacements $1\!\to\!2, 2\!\to\!3, 3\!\to\!4, 4\!\to\!1$ (Fig. \ref{fig-1}(c)). For the discretization of $H_{\rm ME}$, only quadratic terms such as $({\bf s}_i\cdot{\bf s}_j)^2$ in the expansion of $\left(g^{ab}\frac{\partial{\bf s}}{\partial x^{a}}\cdot\frac{\partial{\bf s}}{\partial x^{b}}\right)^2$ are used. 

In these discrete calculations, we extract an intensive part of the energies, the so-called position- and direction-dependent coupling constants 
\begin{eqnarray}
	\label{Gamma-X}
  \Gamma_{ij}=\hat{\Gamma}^{-1}\gamma_{ij}, \quad \hat{\Gamma}=\bar{n}\, \langle \gamma^{\rm{iso}}\rangle, 
\qquad ({\rm FMI,DMI}),\\
	\label{Gammas}
	\begin{split}
		& \gamma_{12}=\frac{v_{12}}{v_{13}v_{14}}+\frac{v_{21}}{v_{23}v_{24}},\quad     \gamma_{13}=\frac{v_{13}}{v_{12}v_{14}}+\frac{v_{31}}{v_{32}v_{34}}, \\
		& \gamma_{14}=\frac{v_{14}}{v_{13}v_{14}}+\frac{v_{41}}{v_{42}v_{43}},\quad  
		 \gamma_{23}=\frac{v_{23}}{v_{21}v_{24}}+\frac{v_{32}}{v_{31}v_{34}}, \\
		& \gamma_{24}=\frac{v_{24}}{v_{21}v_{23}}+\frac{v_{42}}{v_{41}v_{43}},\quad    \gamma_{34}=\frac{v_{34}}{v_{31}v_{32}}+\frac{v_{43}}{v_{41}v_{42}}
	\end{split}
\end{eqnarray}
for $H_{\rm FM}$ and $H_{\rm DM}$, and 
\begin{eqnarray}
	\label{Gamma-X-ME}
\Omega_{ij}=\hat{\Omega}^{-1}\omega_{ij}, \quad \hat{\Omega}=\bar{n}\, \langle \omega^{\rm{iso}}\rangle, 
\qquad ({\rm MEC}),\\
	\label{Gammas-ME}
	\begin{split}
	& \omega_{12}=\frac{v_{12}^3}{v_{13}v_{14}}+\frac{v_{21}^3}{v_{23}v_{24}},\quad     \omega_{13}=\frac{v_{13}^3}{v_{12}v_{14}}+\frac{v_{31}^3}{v_{32}v_{34}}, \\
	& \omega_{14}=\frac{v_{14}^3}{v_{13}v_{14}}+\frac{v_{41}^3}{v_{42}v_{43}},\quad   \omega_{23}=\frac{v_{23}^3}{v_{21}v_{24}}+\frac{v_{32}^3}{v_{31}v_{34}}, \\
	& \omega_{24}=\frac{v_{24}^3}{v_{21}v_{23}}+\frac{v_{42}^3}{v_{41}v_{43}},\quad  
	  \omega_{34}=\frac{v_{34}^3}{v_{31}v_{32}}+\frac{v_{43}^3}{v_{41}v_{42}} 
\end{split}
\end{eqnarray}
for $H_{\rm ME}$. The difference between $\gamma_{ij}$ and $\omega_{ij}$ comes from the quadratic form of the integrand of $H_{\rm ME}$ in Eq. (\ref{cont_FDM}).

Here, we describe only $\Gamma_{ij}$ for simplicity; the definitions of these two are the same except for the expressions of $\gamma_{ij}$ and $\omega_{ij}$ and their $v_{ij}$ in Eqs. (\ref{Finsler-unit-length-FDMI}) and (\ref{Finsler-unit-length-ME}). We should note that $\hat{\Gamma}$ in  $\Gamma_{ij}$ is an irrelevant normalization constant, and a non-trivial interaction comes from $\gamma_{ij}$ in Eq. (\ref{Gammas}). The position- and direction-dependence of interaction is effectively introduced as an intensive part of energy in Hamiltonian via $\Gamma_{ij}$ in the FG modeling.

In Eq. (\ref{Gamma-X}), $\bar{n} (=\!\frac{\sum_{\Delta(ij)}1}{\sum_{ij}1})$ denotes the mean value of $n_{ij}\!=\!\sum_{\Delta(ij)}1$, which is the total number of tetrahedrons $\Delta$ sharing bond $ij$ (Fig. \ref{fig-2}(c)), and $\langle \gamma^{\rm{iso}}\rangle\!=\!\langle\frac{\sum_{\Delta}\sum_{ij(\Delta)} \gamma_{ij}}{\sum_{\Delta}\sum_{ij(\Delta)}1}\rangle$ denotes the ensemble average calculated with 1000 isotropic configurations of $\vec{\tau}$. Note that the random configurations for $\langle \gamma^{\rm{iso}}\rangle$ in $\Omega_{ij}$ include a randomly distributed $\zeta$ on the bonds. The reason for introducing the $v_0$-dependent factor $\langle \gamma^{\rm{iso}}\rangle$ is that $\gamma_{ij}$ varies depending on $v_0$ in Eqs. (\ref{Finsler-unit-length-FDMI}) and (\ref{Finsler-unit-length-ME}), and the parameters $\lambda$, $D$ and $\alpha$ should be changed according to any variations in $v_0$ if $\langle \gamma^{\rm{iso}}\rangle$ is not included.

We briefly show the surface effect for GC. There are two possible sources; static and dynamical, for the surface effect. The first one comes from the lattice structure and the other from a dynamical effect in FG modeling. The first one, which is the static part, is shown using the coupling constant in Eq. (\ref{Gamma-X}) as follows. It is noteworthy that the lattice average $\overline{\Gamma}\!=\!\frac{\sum_{\Delta}\sum_{ij(\Delta)} \Gamma_{ij}}{\sum_{\Delta}\sum_{ij(\Delta)}}$  of $\Gamma_{ij}$  for an isotropic configuration of  $\vec{\tau}$ satisfy $\overline{\Gamma}(=\!\overline{\Gamma}_{ij})\!\simeq\!\bar{n}^{-1}$. Indeed, let $\gamma_{ij}$ be a coefficient obtained with an isotropic configuration of  $\vec{\tau}$ and $\overline{\gamma}$ be the lattice average of  $\gamma_{ij}$. Then, we have  that $\overline{\Gamma}\!=\!\bar{n}^{-1}\langle \gamma^{\rm{iso}}\rangle^{-1}\overline{\gamma}\!=\!\frac{1}{\bar{n}\langle\gamma^{\rm{iso}}\rangle}\frac{\sum_{\Delta}\sum_{ij(\Delta)}\gamma_{ij}}{\sum_{\Delta}\sum_{ij(\Delta)}1}\!\simeq\!\bar{n}^{-1}$ because the lattice average $\overline{\gamma}(\!=\!\frac{\sum_{\Delta}\sum_{ij(\Delta)} \gamma_{ij}}{\sum_{\Delta}\sum_{ij(\Delta)}1})$ for isotropic $\vec{\tau}$  is close to $\langle\gamma^{\rm{iso}}\rangle$. Using this $\overline{\Gamma}$, the Hamiltonian of DMI in Eq. (\ref{Hamiltonians}) for isotropic $\vec{\tau}$ can be written as 
\begin{eqnarray}
\begin{split}
H_{\rm DM}=&\sum_{\Delta}\sum_{ij(\Delta)}\Gamma_{ij}\vec{e}_{ij}\cdot {\bf s}_i\times{\bf s}_j\\ 
\to&\sum_{ij}(\bar{n}^{-1}\sum_{\Delta(ij)}1)\vec{e}_{ij}\cdot {\bf s}_i\times{\bf s}_j=\sum_{ij}(n_{ij}/\bar{n})\vec{e}_{ij}\cdot {\bf s}_i\times{\bf s}_j,
\end{split}
\end{eqnarray} 
where $\Gamma_{ij}$ is replaced by $\overline{\Gamma}\!\simeq\!\bar{n}^{-1}$ and the summation convention $\sum_{ij}\sum_{\Delta(ij)}\!=\!\sum_{\Delta}\sum_{ij(\Delta)}$ is used. The replacement of $\Gamma_{ij}$ with a constant $\overline{\Gamma}$ is  to neglect all dynamical effects implemented in FG modeling and to leave only contribution from the lattice structure. This form of $H_{\rm DM}\!=\!\sum_{ij}(n_{ij}/\bar{n})\vec{e}_{ij}\cdot {\bf s}_i\times{\bf s}_j$ includes the weight $n_{ij}/\bar{n}$, which modifies the interaction $\vec{e}_{ij}\cdot {\bf s}_i\times{\bf s}_j$ to be dependent on the position of bond $ij$ whether it is on the surface or inside. The number $n_{ij}\!=\!\sum_{\Delta(ij)}1$ is smaller (larger) than the mean value $\bar{n}$ on the surface (bulk) (Table \ref{table-2} and Fig. \ref{fig-2}(c)). Thus, we confirm that this weight difference is one possible origin of the surface effect in the small strain region $f\!\to\!0$, where $\vec{\tau}$ is isotropic. This GC effect at $f\!=\!0$ is confirmed in Section \ref{size_effect}.

When an anisotropic configuration of $\vec{\tau}$ (and $\zeta$) is reflected in $\gamma_{ij}$, the weight difference is further enlarged in the region of $f\!>\!0$ in the incomplete skyrmion phase (Fig. \ref{fig-9}(a)). This enlargement caused by anisotropic $\vec{\tau}$ is the dynamical part of the surface effect for GC. Here, we should emphasize that the surface effect is the difference in the interaction weight in $H_{\rm DM}$; effective DMI coupling on the side surface is considerably smaller than the bulk one (Fig. \ref{fig-9}(a)).

% App C
\section{Decomposition of interaction coefficient and the corresponding energies \label{App-C}}
% f-C-1
\begin{figure}[h]
	\centering{}
	\includegraphics[width=12.5cm,clip]{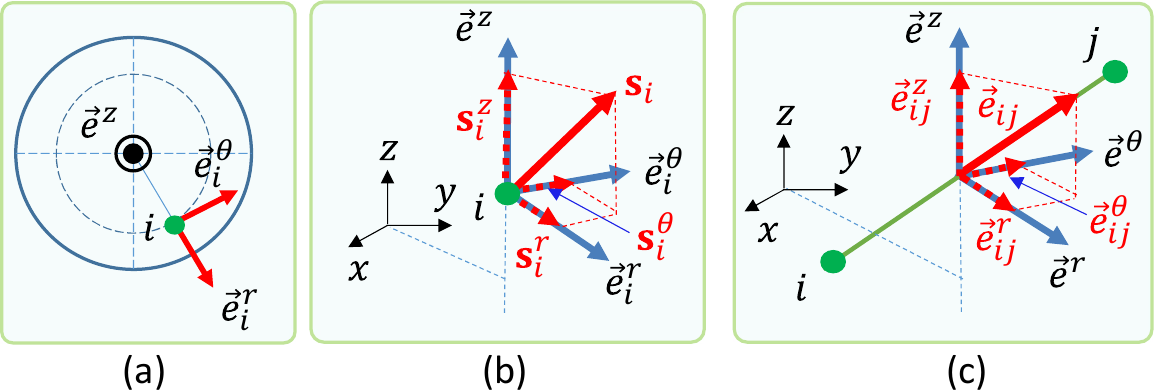}
	\caption{
		(a) Polar coordinate position of vertex $i$. The coordinate origin is at the center of nanodot.   (b) Spin vector ${\bf s}_i$ is decomposed into the radial $\vec{e}_i^{\;r}$, tangential  $\vec{e}_i^{\;\theta}$ and $\vec{e}^{\;z}$ directions at vertex $i$, where  ${\bf s}_{i}^{\mu}\!=\!({\bf s}_{i}\!\cdot\! \vec{e}_i^{\;\mu})\vec{e}_i^{\;\mu}, (\mu\!=\!r,\theta,z)$.  (c)  The unit vector $\vec{e}_{ij}$ from vertices $i$ to $j$ can be decomposed into the $\vec{e}^{\;r}$,  $\vec{e}^{\;\theta}$ and $\vec{e}^{\;z}$ components, where $\vec{e}^{\;r}$ and $\vec{e}^{\;\theta}$ are given by Eq. (\ref{unit-vectors-ij}) and $\vec{e}_{ij}^{\;{\mu}}\!=\!(\vec{e}_{ij}\!\cdot\! \vec{e}^{\;\mu})\vec{e}^{\;\mu}, (\mu\!=\!r,\theta,z)$. 	\label{fig-C-1} }
\end{figure}
Let  $\vec{e}_{ij}$ be the unit vector from vertices $i$ to $j$ and $\vec{e}_i^{\;r}$, $\vec{e}_i^{\;\theta}$ and $\vec{e}^{\;z}$ be the unit vectors along the $r$, $\theta$ and $z$ directions at the vertex $i$ as illustrated in Figs. \ref{fig-C-1}(a),(b), where the coordinate origin is at the center of nanodot and $\vec{e}_{i}^{\; z}$ is independent of $i$ and written as $ \vec{e}^{\; z}$. Here, we define unit vectors
\begin{eqnarray}
	\label{unit-vectors-ij}
				\vec{e}^{\; \mu}=\left(\vec{e}_i^{\;\mu}+\vec{e}_j^{\;\mu}\right)/\|\vec{e}_i^{\;\mu}+\vec{e}_j^{\;\mu}\ \|, \quad(\mu=r,\theta)
\end{eqnarray}
to represent the $r$ and $\theta$ directions at the center of bond $ij$ (Fig. \ref{fig-C-1}(c)). 
Then, we have a decomposition of $\vec{e}_{ij}$ such that $\vec{e}_{ij}\!=\!(\vec{e}_{ij}\cdot\vec{e}^{\;r})\vec{e}^{\;r}\!+\!(\vec{e}_{ij}\cdot\vec{e}^{\;\theta})\vec{e}^{\;\theta}\!+\!(\vec{e}_{ij}\cdot\vec{e}^{\;z})\vec{e}^{\;z}$. Since $\Gamma_{ij}$ is defined on bond $ij$ along  $\vec{e}_{ij}$, $\Gamma_{ij}$ is naturally direction-dependent. To extract this direction dependence,  we use the vector  $\Gamma_{ij}\vec{e}_{ij}$, which is decomposed as
\begin{eqnarray}
	\label{Gamma-decomposition}
\begin{split}
	&\Gamma_{ij}\vec{e}_{ij}=\Gamma_{ij}^{r}\vec{e}^{\;r}+\Gamma_{ij}^{\theta}\vec{e}^{\;\theta}+\Gamma_{ij}^{z}\vec{e}^{\;z},\\
	&\Gamma_{ij}^{r}(\vec{\tau})=\Gamma_{ij}(\vec{\tau})(\vec{e}_{ij}\cdot\vec{e}^{\;r}),\; \Gamma_{ij}^{\theta}(\vec{\tau})=\Gamma_{ij}(\vec{\tau})(\vec{e}_{ij}\cdot\vec{e}^{\;\theta}),\;
\Gamma_{ij}^{z}(\vec{\tau})=\Gamma_{ij}(\vec{\tau})(\vec{e}_{ij}\cdot\vec{e}^{\;z}).
\end{split}
\end{eqnarray}
This decomposition of $\Gamma_{ij}\vec{e}_{ij}$ corresponds to the decomposition of DMI vector (Appendix \ref{App-D}) and is considered a physically meaningful decomposition.
Using these "microscopic" direction-dependent factors $\Gamma_{ij}^{\mu}$, the "macroscopic" direction-dependent coefficients are naturally defined by
\begin{eqnarray}
	\label{HDM-decomp-rtz}
	\Gamma^{\mu} =\frac{1}{N_B}\sum_{ij}\sum_{\Delta(ij)}|\Gamma_{ij}^{\mu}(\vec{\tau})|,\quad (\mu=r,\theta,z).
\end{eqnarray}
The meaning of $\Gamma^{\mu}$ is as follows: If $\Gamma^{r}\!>\!\Gamma^{\theta, z}$ for example, we understand from this inequality that  $\Gamma_{ij}$ along bond $ij$ parallel or almost parallel to $\pm\vec{e}^{\;r}$ is larger than $\Gamma_{ij}$ along bond $ij$ parallel or almost parallel to $\pm\vec{e}^{\;\theta, z}$, because the direction of $\vec{e}_{ij}$ is uniformly distributed.

 Using these $\Gamma^{\mu}$, the corresponding  $H_{\rm DM}$ can also be decomposed into direction-dependent energies:
\begin{eqnarray}
	\label{HDMI-decomposition}
	\begin{split}
	&H_{\rm DM}^{r}=\frac{\sum_{\Delta}\sum_{ij(\Delta)}\Gamma_{ij}(\vec{\tau})(\vec{e}_{ij}\cdot\vec{e}^{\;r})^2\vec{e}_{ij}\cdot {\bf s}_i\times{\bf s}_j}{\frac{1}{N_B}\sum_{\Delta}\sum_{ij(\Delta)}|\Gamma_{ij}^{r}|},\\
	&H_{\rm DM}^{\theta}=\frac{\sum_{\Delta}\sum_{ij(\Delta)}\Gamma_{ij}(\vec{\tau})(\vec{e}_{ij}\cdot\vec{e}^{\;\theta})^2\vec{e}_{ij}\cdot {\bf s}_i\times{\bf s}_j}{\frac{1}{N_B}\sum_{\Delta}\sum_{ij(\Delta)}|\Gamma_{ij}^{\theta}|},\\
	&H_{\rm DM}^{z}=\frac{\sum_{\Delta}\sum_{ij(\Delta)}\Gamma_{ij}(\vec{\tau})(\vec{e}_{ij}\cdot\vec{e}^{\;z})^2\vec{e}_{ij}\cdot {\bf s}_i\times{\bf s}_j}{\frac{1}{N_B}\sum_{\Delta}\sum_{ij(\Delta)}|\Gamma_{ij}^{z}|},
	\end{split}
\end{eqnarray}
where $N_B\!=\!\sum_{ij}1$ is the total number of bonds. Using these expressions, we have 
\begin{eqnarray}
\begin{split}
	\label{HDMI-decomposition-2}
	H_{\rm DM}=&\sum_{\Delta}\sum_{ij(\Delta)}\Gamma_{ij}(\vec{\tau})\vec{e}_{ij}\cdot {\bf s}_i\times{\bf s}_j\\
	=&\sum_{\Delta}\sum_{ij(\Delta)}\Gamma_{ij}(\vec{\tau})\left[(\vec{e}_{ij}\cdot\vec{e}^{\;r})^2+(\vec{e}_{ij}\cdot\vec{e}^{\;r})^2+(\vec{e}_{ij}\cdot\vec{e}^{\;r})^2\right]\vec{e}_{ij}\cdot {\bf s}_i\times{\bf s}_j\\
	=&\left(\frac{1}{N_B}\sum_{\Delta}\sum_{ij(\Delta)}|\Gamma_{ij}^{r}(\vec{\tau})|\right) H_{\rm DM}^{r}+\left(\frac{1}{N_B}\sum_{\Delta}\sum_{ij(\Delta)}|\Gamma_{ij}^{\theta}(\vec{\tau})|\right) H_{\rm DM}^{\theta}\\
	&+\left(\frac{1}{N_B}\sum_{\Delta}\sum_{ij(\Delta)}|\Gamma_{ij}^{z}(\vec{\tau})|\right) H_{\rm DM}^{z},
\end{split}
\end{eqnarray}
and therefore, the lattice average is given by
\begin{eqnarray}
	\label{HDM-decomp-rtz}
%	\begin{split}
		H_{\rm DM}=\Gamma^{r} H_{\rm DM}^{r}+\Gamma^{\theta} H_{\rm DM}^{\theta}+\Gamma^{z} H_{\rm DM}^{z}.
%	\end{split}
\end{eqnarray}
 Eq. (\ref{HDM-decomp-rtz}) is satisfied only in the lattice average. However, for the ensemble average  $\langle *\rangle$  calculated via MC simulations, it is reasonable to assume $\Gamma^{\mu}\!\simeq\!\langle \Gamma^{\mu}\rangle$ and $ H_{\rm DM}^{\mu}\!\simeq\!\langle  H_{\rm DM}^{\mu}\rangle$ in the equilibrium configurations, and hence, we use this decomposition of $H_{\rm DM}$ for the simulation data to evaluate the direction-dependent effective coefficients and energies. 
We should note that $\Gamma_{ij}$ in $H_{\rm FM}$  and $\Omega_{ij}$ in $H_{\rm ME}$ in Eq. (\ref{Hamiltonians}) are defined on bond $ij$ and can also be decomposed into $\mu$ components in the same procedure described in Eq. (\ref{Gamma-decomposition}).  The directional dependence of $H_{\rm FM}$ and $H_{\rm ME}$  can also be evaluated with the same procedure for $H_{\rm DM}$.  

For the calculation of $H_{\rm DM}^\mu({\rm side})$ on the side surface (Figs. \ref{fig-9}(d),(e)), the sum $\sum_{ij}\sum_{\Delta(ij)}(=\!\sum_{\Delta}\sum_{ij(\Delta)})$ in the expressions of Eq. (\ref{HDMI-decomposition})  should be limited to the bonds $ij$ on the side surface. In this case, $N_B$ is the total number of bonds on the side surface.

% f-C-2
\begin{figure}[h]
	\centering{}\includegraphics[width=10.5cm,clip]{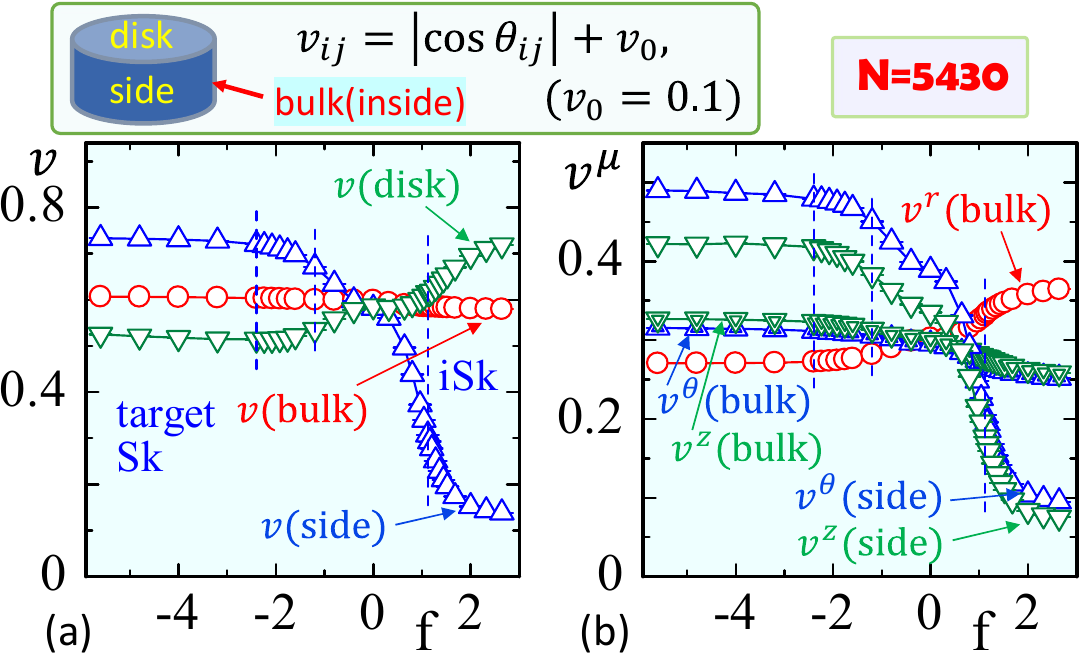}
	\caption{
		(a) Position-dependent  Finsler length $v$ vs. $f$ for the cosine type $v_{ij}\!=\!|\cos \theta_{ij}|\!+\!v_0$ ($v_0\!=\!0.1$) for DMI (and FMI) obtained on the $N\!=\!5430$ lattice, where $\cos\theta_{ij}\!=\!\vec{\tau}\cdot\vec{e}_{ij}$ (Fig. \ref{fig-B-1}(b))  and (b) the direction-dependent ($\Leftrightarrow (r,\theta,z)$-dependent) Finsler length $v^\mu$ vs. $f$. 
		\label{fig-C-2} }
\end{figure}
Finally, in this appendix, we plot the unit Finsler length for DMI in Fig. \ref{fig-C-2}(a) calculated using the lattice average
\begin{eqnarray}
	\label{vij-bonds}
	\begin{split}
		v=\frac{1}{N_B}\dot{\sum}_{ij} \frac{v_{ij}+v_{ji}}{2}
	\end{split}
\end{eqnarray}
on the side surface (side), the upper/lower disk (disk), and the inside (bulk).  The components $v^\mu, (\mu\!=\!r,\theta,z)$ (Fig. \ref{fig-C-2}(b)) are obtained with the absolute components of the vector $\vec{v}_{ij}(=\!\frac{v_{ij}\!+\!v_{ji}}{2}\vec{e}_{ij}$) such that
 \begin{eqnarray}
 	\label{dir-dep-vij}
 	\begin{split}
 		v^\mu=\frac{1}{N_B}\dot{\sum}_{ij} \frac{v_{ij}+v_{ji}}{2}|e_{ij}^\mu|,
 	\end{split}
 \end{eqnarray}
where $N_B=\dot{\sum}_{ij}1$. The meaning of $\dot{\sum}_{ij}$, which is the same as that in Eq. (\ref{energies-per-bond}),  is that the sum over bonds ${\sum}_{ij}$ is limited to the disk, or side, or bulk. The index symmetrization of $v_{ij}$ for $v$ and $v^\mu$ is necessary owing to $v_{ij}\!\not=\!v_{ji}$, in contrast with the case of $\Gamma_{ij}$, where $\Gamma_{ij}\!=\!\Gamma_{ji}$ (Eq. (\ref{Gamma-X})). Without the symmetrization, ${\sum}_{ij} v_{ij}$ depends on the labeling of vertex $i$ and becomes ill-defined because the symbol $\sum_{ij}$ implies the sum over pairs of vertices $i$ and $j$ connected by bond under the condition $i\!>\!j$. The sum over bonds $(1/N_B){\sum}_{ij} \frac{v_{ij}+v_{ji}}{2}$ is the same as  the sum over bonds $(1/2N_B){\sum}_{ij} v_{ij}$ without the condition $i\!>\!j$. Finsler length $v$  is direction independent, whereas $v^\mu$ depends on the direction $\mu(=\!r, \theta, z)$. 

We find from Fig.  \ref{fig-C-2}(a) that  $v({\rm bulk})$ is almost constant and  $v({\rm side})$ and  $v({\rm disk})$ depend on $f$ for positive $f$. In the incomplete skyrmion phase, $v({\rm side})\!\ll\!v({\rm bulk})$, which is considered to be the origin of the surface effect in the FG modeling.  From Fig.  \ref{fig-C-2}(b), we find that $v^r({\rm bulk})$ ($v^{\theta, z}({\rm bulk})$) increases (decrease) with increasing $f$, and hence, the variation of $v^r({\rm bulk})$ is opposite to those of $v^{\theta, z}({\rm bulk})$. The observation that $v^r({\rm bulk})$ increases for $f\!>\!0$ indicates that $v_{ij}$ on bond $ij$ in the direction $\vec{e}_{ij}$ almost parallel to $\vec{e}_{ij}^{\; r}$ is larger than that on bonds $ij$ of $\vec{e}_{ij}$ almost perpendicular to $\vec{e}_{ij}^{\; r}$. It is noteworthy that this behavior of $v^\mu$ is reflected in $\Gamma({\rm bulk})$ and $\Gamma^\mu({\rm bulk})$ in Figs. \ref{fig-9}(a),(c), though $\Gamma_{ij}$ is a rational function of $v_{ij}$ in Eqs. (\ref{Gamma-X}) and (\ref{Gammas}). We should note that in the isotropic case at $f\!\to\!0$, $v_{ij}$ is at random and independent of $v_{ji}$, whereas $v_{ij}\!\simeq\!v_{ji}$ in the anisotropic case $|f|>0$ where $\tau$ is uniformly aligned. We find no position dependence of $v$ on $f$ at $f\!=\!0$. The directional independence of $v$ implies that the original isotropic model ($\Leftrightarrow\! \Gamma_{ij}\!=\!{\rm constant}$ ) is restored in the limit of  $f\!\to\!0$, where $\vec{\tau}$ randomly distributes and effectively has no influence on ${\bf s}$.

% App D
\section{DMI vector \label{App-D}}
 Effective coupling constants $\sum_{\Delta(ij)}\Gamma_{ij}^{\mu}(\vec{\tau}), \;  (\mu=r,\theta,z)$ are understood to be $\mu$ components of a DMI vector. It is interesting to observe that DMI vector is composed of primitive objects in FG modeling. First,  $H_{\rm DM}$ can be written as 
 \begin{eqnarray}
 	\label{HDMI-with-DMI-vector}
 H_{\rm DM}=\sum_{ij}{\bf D}_{ij}\cdot{\bf s}_i\!\times\!{\bf s}_j,
\end{eqnarray}
where ${\bf D}_{ij}$ is a DMI vector and  $\sum_{ij}$ denotes the sum over bonds $ij$. Second,  $H_{\rm DM}$ in Eq. (\ref{Hamiltonians}) can be expressed by the sum over bonds:
\begin{eqnarray}
	\label{HDMI-sum-bonds}
	H_{\rm DM}=\sum_{ij}\sum_{\Delta(ij)}\Gamma_{ij}(\vec{\tau})\vec{e}_{ij}\cdot {\bf s}_i\times{\bf s}_j,
\end{eqnarray}
where $\sum_{\Delta(ij)}$ denotes the sum over tetrahedrons sharing bond $ij$. The expression in Eq. (\ref{HDMI-sum-bonds}) is obtained from   $H_{\rm DM}$  in Eq. (\ref{Hamiltonians}) by using the identity $\sum_{\Delta}\sum_{ij(\Delta)}\!=\!\sum_{ij}\sum_{\Delta(ij)}$ (Fig. \ref{fig-2} and the caption).  Therefore, comparing Eqs. (\ref{HDMI-with-DMI-vector}) and (\ref{HDMI-sum-bonds}), we obtain
\begin{eqnarray}
	\label{DMI-vector}
	\begin{split}
	{\bf D}_{ij}=&\sum_{\Delta(ij)}\Gamma_{ij}(\vec{\tau})\vec{e}_{ij} \\
	            =&\sum_{\Delta(ij)}\Gamma_{ij}(\vec{\tau})\left[(\vec{e}_{ij}\cdot\vec{e}^{\;r})\vec{e}^{\;r}\!+\!(\vec{e}_{ij}\cdot\vec{e}^{\;\theta})\vec{e}^{\;\theta}\!+\!(\vec{e}_{ij}\cdot\vec{e}^{\;z})\vec{e}^{\;z} \right],
	\end{split}
\end{eqnarray}
and the $\mu$ component
\begin{eqnarray}
	\label{DMI-vector-component}
%	\begin{split}
		{D}_{ij}^\mu={\bf D}_{ij}\!\cdot\!\vec{e}^{\; \mu}
		=\left(\sum_{\Delta(ij)}\Gamma_{ij}(\vec{\tau})\right)\vec{e}_{ij}\cdot \vec{e}^{\; \mu}, \;  (\mu=r,\theta,z),
%	\end{split}
\end{eqnarray}
where $\vec{e}^{\;\mu}$ represents $\vec{e}^{\;r}$, $\vec{e}^{\;\theta}$ and $\vec{e}^{\;z}$ (see Eq. (\ref{unit-vectors-ij}) and Fig. \ref{fig-C-1}(c)). Since $\vec{e}_{ij}\cdot \vec{e}^{\; \mu}$ in Eq. (\ref{DMI-vector-component}) is independent of the sum $\sum_{\Delta(ij)}$ and $\Gamma_{ij}(\vec{\tau})\!>\!0$, we find 
$|{D}_{ij}^\mu|=\sum_{\Delta(ij)}\Gamma_{ij}(\vec{\tau})|\vec{e}_{ij}\cdot \vec{e}^{\; \mu}|$. If we write the lattice average of $|{D}_{ij}^\mu|$  by $D^\mu$ such that $D^\mu\!=\!\frac{1}{N_B}\sum_{ij}|D_{ij}^{\mu}(\vec{\tau})|\!=\!\frac{1}{N_B}\sum_{ij}\sum_{\Delta(ij)}\Gamma_{ij}(\vec{\tau})|\vec{e}_{ij}\cdot \vec{e}^{\; \mu}|$, then $\Gamma^\mu(=\!\frac{1}{N_B}\sum_{\Delta}\sum_{ij(\Delta)}|\Gamma_{ij}^{\mu}(\vec{\tau})|)$ is identified with $D^\mu$:
\begin{eqnarray}
	\label{mean-DMI-vector-component}
	\begin{split}
{D}^\mu			=&\frac{1}{N_B}\sum_{ij}\sum_{\Delta(ij)}\Gamma_{ij}(\vec{\tau})|\vec{e}_{ij}\cdot \vec{e}^{\; \mu}|\\
		=&\frac{1}{N_B}\sum_{\Delta}\sum_{ij(\Delta)}\Gamma_{ij}(\vec{\tau})|\vec{e}_{ij}\cdot \vec{e}^{\; \mu}|\\
		=&\Gamma^{\mu}, \;  (\mu=r,\theta,z).
	\end{split}
\end{eqnarray}
 Using these relations, Eq. (\ref{HDM-decomp-rtz}) can also be written as 
  \begin{eqnarray}
	\label{micro-macro-correspondence}
	 H_{\rm DM}=
 D^{r}  H_{\rm DM}^{r}+ D^{\theta} H_{\rm DM}^{\theta}+ D^{z} H_{\rm DM}^{z}.
 \end{eqnarray}
  Therefore, the extensive parts $ H_{\rm DM}^{\mu}, (\mu\!=\!r,\theta,z)$ are considered to be physically meaningful direction-dependent energies, expressed to be determined as a response to the intensive parts $D^{\mu}$, which are also dynamically determined. 
  
   The symbol $\Gamma$ is used for the lattice average defined by
 \begin{eqnarray}
	\label{mean-DMI-coefficient-APP}
 	\Gamma=\frac{1}{N_B}\sum_{ij}\sum_{\Delta(ij)}\Gamma_{ij}(\vec{\tau}),
 \end{eqnarray}
 which corresponds to the lattice average of $D_{ij}\!=\!\|{\bf D}_{ij}\|\!=\!\sum_{\Delta(ij)}\Gamma_{ij}(\vec{\tau})$.
 
%-------------------------------
%\section*{References}
%-------------------------------

%\nocite{*}
%\bibliography{aipsamp}

\begin{thebibliography}{9}



% applications
% write delete
\bibitem{Romming-etal-Science2013} N. Romming, C. Hanneken, M. Menzel, J. E. Bickel, B. Wolter,
K. von Bergmann, A. Kubetzka and R. Wiesendanger, {\it Writing and deleting single magnetic skyrmions}, Science \textbf{341} (6146), 636-639 (2013),  https://doi.org/10.1126/science.1240573.

% Reviews

\bibitem{Fert-etal-NatReview2017} 
A. Fert, N. Reyren and V. Cros, {\it Magnetic skyrmions: advances in physics and potential applications}, Nature Reviews \textbf{2},
17031 (2017), https://doi.org/10.1038/natrevmats.2017.31.

\bibitem{Zhang-JPCM2020} 
X. Zhang, Y. Zhou, K. M. Song, T.-E. Park, J. Xia,
M. Ezawa, X. Liu, W. Zhao, G. Zhao and S. Woo, {\it Skyrmion-electronics: writing, deleting,
	reading and processing magnetic skyrmions toward spintronic applications}, J. Phys.: Condens. Matter {\bf 32} 143001  (2020),  https://doi.org/10.1088/1361-648X/ab5488.

\bibitem{Tokura-Kanazawa-ACS2020}
Y. Tokura and N. Kanazawa, {\it Magnetic Skyrmion Materials}, Chem. Rev. {\bf 121}, 5, pp. 2857–2897 (2021), https://doi.org/10.1021/acs.chemrev.0c00297.

\bibitem{Gobel-etal-PhysRep2021}
B$\ddot{\rm o}$rge G$\ddot{\rm o}$bel, I. Mertig, O. A. Tretiakov, {\it Beyond skyrmions: Review and perspectives of alternative magnetic quasiparticles}, Phys. Rep. {\bf 895} 1-28,  (2021), https://doi.org/10.1016/j.physrep.2020.10.001.

\bibitem{Wang-etal-JMMM2022}
K. Wang, V. Bheemarasetty, J. Duan, S. Zhou and G. Xiao,  {\it Fundamental physics and applications of skyrmions: A review}, J. Mag. Mag. Mat. {\bf 563}, 169905 (2022), https://doi.org/10.1016/j.jmmm.2022.169905.

\bibitem{Li-etal-Wiley2022}
Sheng Li, Xuewen Wang, Theo Rasing,  {\it Magnetic skyrmions: Basic properties and potential 	applications}, Interdisciplinary Materials {\bf 563}, 169905 (2022), https://doi.org/10.1002/idm2.12072.

%observation
% Lorentz-microscopy

\bibitem{Uchida-etal-SCI2006} M. Uchida, Y. Onose,Y. Matsui and Y. Tokura, {\it Real-Space Observation	of Helical Spin Order}, Science \textbf{311}, pp.359-361 (2006), https://doi.org/10.1126/science.11206.

\bibitem{Pfleiderer-etal-Science2009} 
S. M$\ddot{\rm u}$hlbauer, B. Binz, F. Jonietz, C. Pfleiderer, A. Rosch, A. Neubauer, R. Georgii, P. B$\ddot{\rm o}$ni, {\it Skyrmion Lattice in a Chiral Magnet}, Science,  {\bf 123}, 915 (2009),  https://doi.org/10.1126/science.116676.

% 2d system TEM & neutron  
\bibitem{Yu-etal-Nature2010} X. Yu, Y. Onose, N. Kanazawa, J.H. Park, J.H. Han, Y. Matsui, N. Nagaosa, and Y. Tokura, {\it Real-space observation of a two-dimensional skyrmion crystal}, Nature \textbf{465}, 901-904 (2010), https://doi.org/10.1038/nature09124.


% magetostriction
%-----------------------------------------------
\bibitem{Bogdanov-PRL2001} 
A. N. Bogdanov, and U. K. R${\ddot {\rm o}}{\rm \beta}$ler, {\it Chiral Symmetry Breaking in Magnetic Thin Films and Multilayers}, Phys. Rev. Lett., \textbf{87}, 037203 (2001), https://doi.org/10.1103/PhysRevLett.87.037203.

\bibitem{Butenko-etal-PRB2010} 
A. B. Butenko, A. A. Leonov, U. K. R${\ddot {\rm o}}{\rm \beta}$ler, and A. N. Bogdanov, {\it Stabilization of skyrmion textures by uniaxial distortions in noncentrosymmetric cubic helimagnets}, Phys. Rev. B \textbf{82}, 052403 (2010), https://doi.org/10.1103/PhysRevB.82.052403.


% theory geom. confinement
\bibitem{Rohart-Thiaville-PRB2013} 
S. Rohart and A. Thiaville, {\it Skyrmion confinement in ultrathin film nanostructures in the presence of Dzyaloshinskii-Moriya interaction}, Phys. Rev. B {\bf 88}, 184422 (2013), 
https://doi.org/10.1103/PhysRevB.88.184422.

% experiment geom. confinement
% FeGe naostripe
\bibitem{HDu-etal-NatCom2015} 
H. Du, R. Che, L. Kong X. Zhao, C. Jin, C. Wang, J. Yang, W. Ning, R. Li, C. Jin, X. Chen, J. Zang, Y. Zhang, and M. Tian, {\it Edge-mediated skyrmion chain and its collective dynamics in a confined geometry}, Nature Comm. {\bf 6}, 8504 (2015), https://doi.org/10.1038/ncomms9504.

% confinements
\bibitem{Matsumoto-etal-NanoLett2018} 
T. Matsumoto, Y.-G. So, Y. Kohno, Y. Ikuhara, and N. Shibata, {\it Stable Magnetic Skyrmion States at Room Temperature Confined to Corrals of Artificial Surface Pits Fabricated by a Focused Electron Beam}, Nano Lett. {18},  754-762 (2018),  https://doi.org/10.1021/acs.nanolett.7b03967.


% nanostripe, Simulation, frustrated spins at edge
\bibitem{CJin-etal-NatCom2017} 
C. Jin, Zi-An Li, A. Kov${\rm\acute{a}}$cs, J. Caron, F. Zheng, F. N. Rybakov, N. S. Kiselev, H. Du, S. Bl${\rm\ddot{u}}$gel, M. Tian, Y. Z., M. Farle, and Rafal E. Dunin-Borkowski, {\it Control of morphology and formation of highly geometrically confined magnetic skyrmions}, Nature Comm. {\bf 8}, 15569 (2017), https://doi.org/10.1038/ncomms15569.

% Sky bubble, dipole-diploe int.
\bibitem{ZHou-etal-AcsNano2019} 
Z. Hou, Q. Zhang, G. Xu, S. Zhang, C. Gong, B. Ding, H. Li, F. Xu, Y. Yao, E. Liu, G. Wu, X. Zhang, and W. Wa, {\it Manipulating the Topology of Nanoscale Skyrmion Bubbles by Spatially Geometric Confinement}, ACS Nano. {\bf 13}, 922-929 (2019), https://doi.org/10.1021/acsnano.8b09689.

% multilayered & zero mag.f. nanodots
\bibitem{PHo-etal-PRAp2019} 
Pin Ho, Anthony K.C. Tan, S. Goolaup, A.L. Gonzalez Oyarce, M. Raju, L.S. Huang, Anjan Soumyanarayanan, and C. Panagopoulos, {\it Geometrically Tailored Skyrmions at Zero Magnetic Field in Multilayered Nanostructures}, {Phys. Rev. Appl.} {\bf 11}, 024064 (2019), 
https://doi.org/10.1103/PhysRevApplied.11.024064.


% stabilazation by uniaxial strain 
\bibitem{Seki-etal-PRB2017}
S. Seki, Y. Okamura, K. Shibata, R. Takagi, N. D. Khanh, F. Kagawa, T. Arima, and Y. Tokura, {\it Stabilization of magnetic skyrmions by uniaxial tensile strain}, Phys.Rev. B \textbf{96}, 220404(R) (2017), https://doi.org/10.1103/PhysRevB.96.220404.


% nanodots under tensile strain
\bibitem{YWang-etal-NatCom2020} 
Y. Wang, L. Wang, J. Xia, Z. Lai, G. T. X. Zhang, Z. Hou, X. Gao, W. Mi, C. Feng, M. Zeng, G. Zhou, G. Yu, G. Wu, Y. Zhou, W. Wang, X. Zhang, and J. Liu, {\it Electric-field-driven non-volatile multi-state switching of individual skyrmions in a multiferroic heterostructure},
Nature Comm. {\bf 11}, 3577 (2020), https://doi.org/10.1038/s41467-020-17354-7.



% nanodisk
\bibitem{Li-etal-PRAP2020} 
H. Li, C. A. Akosa, P. Yan, Y. Wang, and Z. Cheng, {\it Stabilization of Skyrmions in a Nanodisk Without an External Magnetic Field }, Phys. Rev. Applied. {\bf 13}, 034046 (2020), https://doi.org/10.1038/s41467-020-17354-7.



% target skyrmions, incomplete skyrmions
\bibitem{Leonov-etal-EPJConf2014}
A. O. Leonov, U. K. R$\ddot{\rm o}$ßler, and M. Mostovoy,  {\it Target-skyrmions and skyrmion clusters in nanowires of chiral magnets}, EPJ Web of Conferences, {\bf 75}, 05002 (2014), https://doi.org/10.1051/jepjconf/20147505002.

\bibitem{Du-etal-PRBL2013}
H. Du, W. Ning, M. Tian and Y. Zhang,  {\it Field-driven evolution of chiral spin textures in a thin helimagnet nanodisk}, Phys. Rev. B {\bf 87}, 014401 (2013);, https://doi.org/10.1103/PhysRevB.87.014401.


\bibitem{Zheng-etal-PRL2017}
F, Zheng, H. Li, S. Wang, D. Song, C. Jin, W. Wei, A. Kov$\acute{\rm a}$cs,
J. Zang, M. Tian, Y. Zhang, H. Du and R. E. Dunin-Borkowski,  {\it Direct Imaging of a Zero-Field Target Skyrmion and Its Polarity Switch
	in a Chiral Magnetic Nanodisk}, Phys. Rev. Lett. {\bf 119}, 197205 (2017), https://doi.org/10.1103/PhysRevLett.119.197205.


\bibitem{Kent-etal-APL2019}
N. Kent, R. Streubel, C. -H. Lambert, A. Ceballos, S.-G. Je, S. Dhuey, Mi-Y. Im, F. B$\ddot{\rm u}$uttner, F. Hellman, S. Salahuddin and P. Fischer,  {\it Generation and stability of structurally imprinted target skyrmions in magnetic multilayers}, Appl. Phys. Lett. {\bf 115}, 112404 (2019), https://doi.org/10.1063/1.5099991.


\bibitem{Beg-etal-SRep2015}
M. Beg, R. Carey, W. Wang, D. Cort$\acute{\rm e}$s-Ortu$\tilde{\rm n}$o, M. Vousden, M.-A. Bisotti, M. Albert, D. Chernyshenko, O. Hovorka, R. L. Stamps and H. Fangohr,  {\it Ground state search, hysteretic behaviour, and reversal mechanism
	of skyrmionic textures in confined helimagnetic nanostructures}, Scientific Reports {\bf 5}, 17137 (2015), https://doi.org/10.1038/srep17137.

\bibitem{Beg-etal-PRB2017}
M. Beg, M. Albert, M. -A. Bisotti, D. Cort$\acute{\rm e}$s-Ortu$\tilde{\rm n}$o, W. Wang, R. Carey, M. Vousden, O. Hovorka, C. Ciccarelli, C. S. Spencer, C. H. Marrows and H. Fangohr,  {\it Dynamics of skyrmionic states in confined helimagnetic nanostructures}, Phys. Rev. B {\bf 95}, 014433  (2017), https://doi.org/10.1103/PhysRevB.95.014433.









% GC on nanodisk
\bibitem{Mehmood-etal-JMMM2022} 
N. Mehmood, J. Wang, C. Zhang, Z. Zeng, J. Wang, and Q. Liu, {\it Magnetic skyrmion shape manipulation by perpendicular magnetic anisotropy excitation within geometrically confined nanostructures}, J. Mag. Mag. Mat. {\bf 545}, 168775 (2022), https://doi.org/10.1016/j.jmmm.2021.168775.

% nanoparticle
\bibitem{Niitsu-etal-Natmatt2022} 
K. Niitsu, Y. Liu, A. C. Booth, X. Yu, N. Mathur, M. J. Stolt, D. Shindo, S. Jin, J. Zang, N. Nagaosa, and Y. Tokura, {\it Geometrically stabilized skyrmionic vortex in FeGe tetrahedral nanoparticles}, {Nature Materials} {\bf 21}, 305-310 (2022), 
https://doi.org/10.1038/s41563-021-01186-x.




% FG modeling conference paper
\bibitem{Diguet-etal-ICFD2023} 
G. Diguet, B. Ducharne, S. El Hog, F. Kato, H. Koibuchi, T. Uchimoto, and H. T. Diep, {\it Finsler Geometry Modeling and Monte Carlo Study on Geometrically Confined skyrmions in Nanodot}, 
{Proc. ICFD2023}, http://arxiv.org/abs/2308.16199.




% Voronoi lattice construction
\bibitem{Friedberg-Ren-NPB1984} 
R. Friedberg and H.-C. Ren, {\it Field theory on a computationally constructed random lattice}, Nucl. Phys. B \textbf{235}, 310-320 (1984), https://doi.org/10.1016/0550-3213(84)90501-7.
%-----------------------------------------------


% MC technique
\bibitem{Metropolis-JCP-1953} 
N. Metropolis, A.W. Rosenbluth, M.N. Rosenbluth, A.H. Teller, and E. Teller, {\it Equation of State Calculations by Fast Computing Machines}, J. Chem. Phys. \textbf{21}, 1087 (1953), https://doi.org/10.1063/1.1699114.

\bibitem{Landau-PRB1976} 
D.P. Landau, {\it Finite-size behavior of the Ising square lattice}, Phys. Rev. B \textbf{13}, 2997 (1976), https://doi.org/10.1103/PhysRevB.13.2997.

% GC model simulation
\bibitem{Diguet-etal-JMMM2023} 
G. Diguet, B. Ducharne, S. El Hog, F. Kato, H. Koibuchi, T. Uchimoto, and H.T. Diep, {\it Monte Carlo studies of skyrmion stabilization under geometric confinement and uniaxial strain}, J. Mag. Mag. Mat. {\bf 579}, 170819 (2023), https://doi.org/10.1016/j.jmmm.2023.170819.


% nonpolar order parameters
\bibitem{Andrienko-JML2018} 
D. Andrienko,  {\it Introduction to liquid crystals}, J. Mol. Liq., {\bf 267}, pp. 520-541 (2018), https://doi.org/10.1016/j.molliq.2018.01.175.



% vortex
\bibitem{Shinjo-etal-science2000} 
T. Shinjo T. Okuno R. Hassdorf, K. Shigeto and T. Ono, {\it Magnetic Vortex Core Observation in Circular Dots of Permalloy}, Science {\bf 289}, 930 (2000), 
https://doi.org/10.1126/science.289.5481.930.

\bibitem{Okuno-etal-JMMM2002} 
T. Okuno, K. Shigeto, T. Ono, K. Mibu and T. Shinjo, {\it MFM study of magnetic vortex cores in circular permalloy dots: behavior in external field}, J. Mag. Mag. Mat. {\bf 240}, pp. 1-6 (2002), 
https://doi.org/10.1016/S0304-8853(01)00708-9.


\end{thebibliography}
% Produces the bibliography via BibTeX.

\end{document}